\documentclass[10pt,journal,compsoc]{IEEEtran}

\ifCLASSOPTIONcompsoc
  \usepackage[compress]{cite}
\else
  \usepackage{cite}
\fi
\usepackage{algorithm}
\usepackage{algpseudocode}
\usepackage{verbatim}
\usepackage{amsmath}
\usepackage{amssymb}
\usepackage{graphicx}
\usepackage{bbm}
\usepackage{listings}
\usepackage{ragged2e}
\usepackage{caption}
\newtheorem{theorem}{Theorem}[section] 
\newtheorem{lemma}[theorem]{Lemma}

\newenvironment{proof}[1][Proof]{\begin{trivlist}
		\item[\hskip \labelsep {\bfseries #1}]}{\end{trivlist}}
\newenvironment{definition}[1][Definition]{\begin{trivlist}
		\item[\hskip \labelsep {\bfseries #1}]}{\end{trivlist}}

\newcommand{\bA}{\mathbf{A}}
\newcommand{\bH}{\mathbf{H}}
\newcommand{\bX}{\mathbf{X}}
\newcommand{\bP}{\mathbf{P}}
\newcommand{\bB}{\mathbf{B}}
\newcommand{\bD}{\mathbf{D}}
\newcommand{\bS}{\mathbf{S}}
\newcommand{\bR}{\mathbf{R}}
\newcommand{\bJ}{\mathbf{J}}
\newcommand{\bGamma}{\mathbf{\Gamma}}
\newcommand{\bPsi}{\mathbf{\Psi}}
\newcommand{\bLambda}{\mathbf{\Lambda}}

\newcommand{\EE}{\mathbb{E}}           

\begin{document}
\title{Joint Embedding of Graphs}
\author{Shangsi Wang, Jes\'us Arroyo,
        Joshua T.~Vogelstein,
        Carey E.~Priebe
         
\IEEEcompsocitemizethanks{\IEEEcompsocthanksitem Shangsi Wang and Carey Priebe are with the Department of Applied Mathematics and Statistics, Johns Hopkins University. \protect E-mail: swang127@jhu.edu, cep@jhu.edu
\IEEEcompsocthanksitem Jes\'us Arroyo is with the Center for Imaging Science, Johns Hopkins University. \protect E-mail: jesus.arroyo@jhu.edu
\IEEEcompsocthanksitem Joshua T.~Vogelstein is with the Department of Biomedical Engineering,  Institute for Computational Medicine, and Kavli Neuroscience Discovery Institute, Johns Hopkins University \protect E-mail: jovo@jhu.edu
\IEEEcompsocthanksitem Shangsi Wang and Jes\'us Arroyo equally contributed to the project.
\IEEEcompsocthanksitem The authors gratefully acknowledge support from  the GRAPHS program of the Defense Advanced Research
Projects Agency (DARPA) contract N66001-14-1-4028, the DARPA SIMPLEX program contract N66001-15-C-4041, the DARPA D3M program administered through AFRL contract FA8750-17-2-0112.
and the DARPA MAA program contract number FA8750-18-2-0066. The U.S. Government is authorized to reproduce and distribute reprints for Governmental purposes notwithstanding any copyright notation thereon. The views and conclusions contained herein are those of the authors and should not be interpreted as necessarily representing the official policies or endorsements, either expressed or implied, of the Air Force Research Laboratory and DARPA or the U.S. Government.}}


\IEEEtitleabstractindextext{
\justify \begin{abstract} 	
Feature extraction and dimension reduction for networks is critical in a wide variety of domains. Efficiently and accurately learning features for multiple graphs has important applications in statistical inference on graphs. We propose a method to jointly embed multiple undirected graphs. Given a set of graphs, the joint embedding method identifies a linear subspace spanned by rank one symmetric matrices and projects adjacency matrices of graphs into this subspace. The projection coefficients can be treated as features of the graphs, while the embedding components can represent vertex features.  We also propose a random graph model for multiple graphs that generalizes other classical models for graphs. We show through theory and numerical experiments that under the model, the joint embedding method produces estimates of parameters with small errors. Via simulation experiments, we demonstrate that the joint embedding method produces features which lead to state of the art performance in classifying graphs. Applying the joint embedding method to human brain graphs, we find it extracts interpretable features with good prediction accuracy in different tasks.
\end{abstract}

}

\maketitle

\IEEEdisplaynontitleabstractindextext
\IEEEpeerreviewmaketitle

\IEEEraisesectionheading{\section{Introduction}\label{sec:introduction}}
\noindent \IEEEPARstart{I}n many problems arising in science and engineering, graphs arise naturally as data structure to capture complex relationships between a set of objects. Graphs have been used in various application domains as diverse as social networks \cite{otte2002social}, internet mapping \cite{govindan2000heuristics}, brain connectomics \cite{bullmore2011brain}, political voting networks \cite{ward2011network},  and many others. The graphs are naturally high dimensional objects with complicated topological structure, which makes graph clustering and classification a challenge to traditional machine learning algorithms. Therefore, feature extraction and dimension reduction techniques are helpful in the applications of learning graph data. In this paper, we propose an algorithm to jointly embed multiple graphs into low dimensional space. We demonstrate through theory and experiments that the joint embedding algorithm produces features which lead to state of the art performance for subsequent inference tasks on graphs.  \\

\noindent There exist a few unsupervised approaches to extract features from graphs. First, classical Principal Component Analysis can be applied by treating each edge of a graph as a raw feature\cite{jolliffe2002principal}. This approach produces features which are linear combinations of edges, but it ignores the topological structure of graphs and the features extracted are not easily interpretable. Second, features can be extracted by computing summary topological and label statistics from graphs\cite{li2011graph,park2013anomaly}. These statistics commonly include number of edges, number of triangles, average clustering coefficient, maximum effective eccentricity, etc. In general, it is hard to know what intrinsic statistics to compute \textit{a priori} and computing some statistics can be computationally expensive. Third, many frequent subgraph mining algorithms are developed \cite{jiang2013survey}. For example, the fast frequent subgraph mining algorithm can identify all connected subgraphs that occur in a large fraction of graphs in a graph data set \cite{huan2003efficient}. Finally, spectral feature selection can also be applied to graphs. It treats each graph as a node and constructs an object graph based on a similarity measure. Features are computed through the spectral decomposition of this object graph \cite{zhao2007spectral}.

\noindent Adjacency Spectral Embedding (ASE) and Laplacian Eigenmap (LE) are proposed to embed a single graph observation \cite{sussman2012consistent, belkin2003laplacian}. The inference task considered in these papers is learning of the block structure of the graph or clustering vertices. Given a set of graphs $\{G_i=(V_i,E_i)\} _{i=1}^{m}$, ASE and LE need to embed an adjacency matrix or Laplacian matrix of $G_i$ individually, and there is no easy way to combine multiple embeddings. The joint embedding method considers the set of graphs together. It takes a matrix factorization approach to extract features for multiple graphs. The algorithm manages to simultaneously identify a set of rank one matrices and project adjacency matrices into the linear subspace spanned by this set of matrices. The joint embedding can be understood as a generalization of ASE for multiple graphs. We demonstrate through simulation experiments that the joint embedding algorithm extracts features which lead to good performance for a variety of inference tasks. In the next section, we review some random graph models and present a model for generating multiple random graphs. In Section 3, we define the joint embedding of graphs and present an algorithm to compute it. In Section 4, we perform some theoretical analyses of our joint embedding. The theoretical results and real data experiments are explored in Section 5. We conclude the paper with a brief discussion of implications and possible future work.

\section{Setting}
We focus on embedding unweighted and undirected graphs for simplicity, although the joint embedding algorithm works on weighted graphs, and directed graphs with some modifications.  Let $\{G_i=(V_i,E_i)\} _{i=1}^{m}$ be $m$ graphs, each with $n$ vertices, and $\bA _i$ be the adjacency matrix of graph $G_i$. The vertices in these graphs should be matched, which means that all the graphs have a common vertex set $V$. The joint embedding algorithm embeds all $G_i$s simultaneously into $\mathbb{R}^d$ and represents $G_i$  by a vector $\lambda_i \in \mathbb{R}^d$. Before discussing the joint embedding algorithm, we need a random graph model on multiple graphs, on which the theoretical analysis is based. Let us first recall a model on a single graph: Random Dot Product Graph \cite{young2007random}. 

\begin{definition}{Random Dot Product Graph (RDPG)}. Let $\mathcal{X}$ be a subset of $\mathbb{R}^d$ such that $x^T y \in [0, 1]$ for all $x, y \in \mathcal{X}$. Let $\bX=[x_1^T,x_2^T,...,x_n^T] \in \mathcal{X}^n$ be a $n\times d$ matrix, and given $\bX$, suppose that $\bA$ is a random $n\times n$ adjacency matrix such that
\[ \bA_{st} \sim Bernoulli(x_s^T x_t). \]
	Alternatively,
	\[ P(\bA|\bX) = \prod_{s<t} (x_s^T x_t) ^{ \bA_{st}} (1-x_s^T x_t)^{1- \bA_{st}}.\]
Also, define $\bP:=\bX\bX^T$ to be edge probability matrix. We write $\bA \sim \operatorname{RDPG}(\bX)$ to denote the distribution of a \emph{random dot product graph} with latent positions $\bX$. When the rows of $\bX$ are not fixed, but instead are random variables with a distribution $F$ on $\mathcal{X}$,  $(\bX,\bA) \sim \operatorname{RDPG}(F)$ denotes the distribution of a \emph{random dot product graph} with latent positions distributed according to $F$.
\end{definition}

The RDPG is a convenient model which is designed to capture the relationship between the vertices of a graph using latent positions, and some extensions have been proposed to capture more general connectivity structures \cite{Rubin-Delanchy2017}. Moreover, other popular models, including the stochastic block model (SBM) \cite{holland1983stochastic} and mixed membership SBM \cite{Airoldi2007}
 are special cases of the RDPG and its generalizations.. The RDPG can be further generalized to a Latent Position Graph by replacing the inner product by a kernel \cite{tang2013universally}. The Adjacency Spectral Embedding of a RDPG adjacency matrix is well studied \cite{sussman2014consistent}. Next, we propose a new random graph model which generalizes the RDPG to multiple graphs.

\begin{definition} Multiple Random Eigen Graphs (MREG). Let $h_1, \ldots, h_d$ be  vectors in $\mathbb{R}^{n}$ with $\|h_i\|_2 = 1, i=1, \ldots,d$, and  denote by $\mathcal{X} \subset \mathbb{R}^d$ the set of vectors satisfying $\sum\limits_{k=1}^{d} \lambda [k] h_k  h_k^T \  \in [0, 1]^{n \times n} $ for all $\lambda \in \mathcal{X}$, where $\lambda[k]$ is the $k$th entry of vector $\lambda$. The random adjacency matrices $\bA_1,\ldots,\bA_m$  follow a $d$-dimensional \emph{multiple random eigen graphs model}, denoted by
\[\{\bA_i\}_{i=1}^m \sim \operatorname{MREG}(\{\lambda_i\}_{i=1}^m,h_1,...,h_d),\]
if the entries of $A_i$ are independent Bernoulli random variables,
	\[ \bA_{i}[s,t] \sim Bernoulli( \sum_{k=1}^{d} \lambda_{i}[k] h_{k} [s] h_{k} [t] ). \]
$\bP_i:=\sum_{k=1}^{d} \lambda_i [k] h_k  h_k^T$ is defined to be the edge probability matrix for graph $i$.  In cases that $\{\lambda_i\}_{i=1}^m$ are of primary interest, they are treated as parameters. When the $\{\lambda_i\}_{i=1}^m$ are random variables,  $F$ denotes their distribution on $\mathcal{X}$, and the model is expressed as
\[\{(\lambda_i,\bA_i)\}_{i=1}^m \sim MREG(F,h_1,...,h_d).\]
\end{definition}

Compared to the RDPG model, MREG is designed to model multiple graphs. The vectors $\{h_k\}_{k=1}^{d}$ are shared across graphs and represent joint latent positions of the vertices, where $(h_j[1],\ldots,h_j[d])\in\mathbb{R}^d$ represents the position of vertex $j$; a  $\lambda_i$ represents the parameter of an individual graph relative to the latent positions $\{h_k\}_{k=1}^{d}$. Note that for a single graph, the edge probability matrix can be written as $\bP_i = \bH\bLambda_i\bH^T$, where $\bH = [h_1 \cdots h_d]\in\mathbb{R}^{n\times d}$ and $\bLambda_i=\text{diag}(\lambda_i)$. When the entries of $\lambda_i$ are non-negative, then $\bX_i = \bH\bLambda_i^{1/2}$ are the latent positions of graph $i$, and hence, on a single graph, RDPG and MREG are equivalent if the edge probability matrix is positive semidefinite. In MREG, we allow self loops to happen. This is mainly for theoretical convenience. Next, we introduce another random graph model: Stochastic Block Model \cite{holland1983stochastic}, which generalizes the Erdos-Renyi (ER) model \cite{erdds1959random} that corresponds to a SBM with only one block. SBM is a widely used model to study the community structure of a  graph \cite{karrer2011stochastic, lyzinski2017community}.\\

\begin{definition} Stochastic Block Model (SBM). Let $\pi$ be a prior probability vector for block membership which lies in the unit $K-1$-simplex. Denote by $\tau=(\tau_1,\tau_2,...,\tau_n) \in [K]^n$ the block membership vector, where $\tau$ is a multinomial sequence with probability vector $\pi$.
	Denote by $\bB \in [0,1]^{K \times K}$ the block connectivity probability matrix. Suppose $\bA$ is a random adjacency matrix given by,
	\[ P(\bA|\tau,\bB)= \prod_{i<j} \bB_{\tau_s,\tau_t}^{\bA_{s,t}} (1-\bB_{\tau_s,\tau_t})^{(1-\bA_{s,t})}\] 
	Then, $\bA$ is an adjacency matrix of a $K$-block stochastic block model graph, and the notation is $\bA \sim SBM(\pi,\bB)$. Sometimes, $\tau$ may also be treated as the parameter of interest, in this case the notation is $\bA \sim SBM(\tau,\bB)$.
\end{definition}

\begin{figure}[!htbp]
	\centering
	\includegraphics[scale=0.6,width=3.0in]{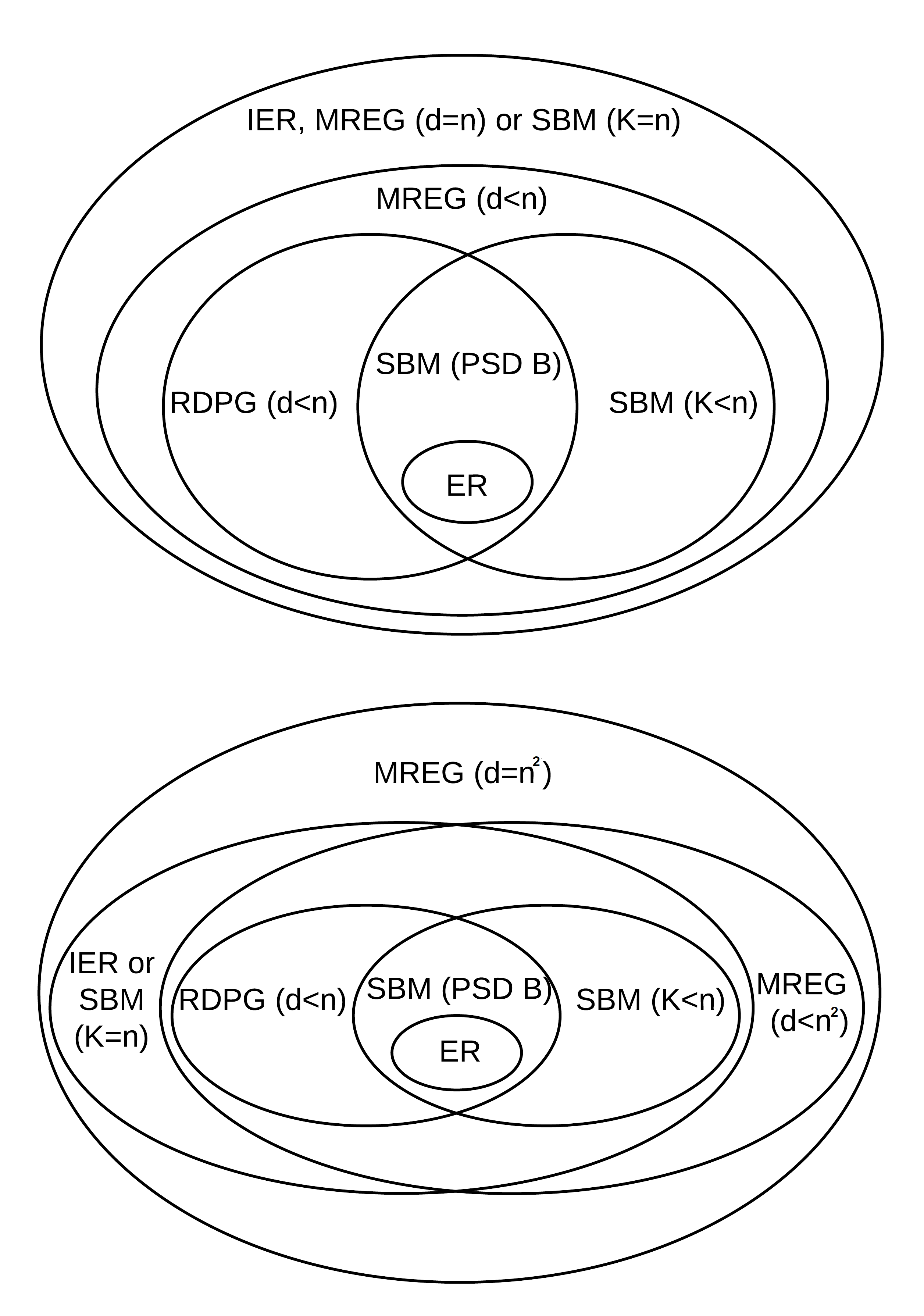}
	\caption{Relationships between random graph models on $1$ graph and multiple graphs. The top panel shows the relationships between the random graph models on $1$ graph. The models considered are those conditioned on latent positions, that is $\tau$, $\bX$ and $\lambda$ in SBM, RDPG and MREG respectively are treated as parameters. ER is a $1$-block SBM. If a graph follows SBM with a positive semidefinite edge probability matrix, it also follows the RDPG model. Any  SBM and  RDPG graph can be represented by a $d$-dimensional MREG model with $d$ being less than or equal to the number of blocks or the dimension of RDPG. On one graph, inhomogeneous ER (IER), $n$-dimensional MREG and $n$-block SBM are equivalent. The bottom panel shows the relationships between the random graph models on multiple graphs. The models considered are those conditioned on latent positions, and for ER, SBM and RDPG graphs are sampled i.i.d. with the same parameters. In this case, MREG has the flexibility to have $\lambda$ differ across graphs, which leads to a more generalized model for multiple graphs.}
	\label{fig:ven}
\end{figure}

The top panel of Figure \ref{fig:ven} shows the relationships between three random graph models defined above and the ER model on $1$ graph. The models considered are those conditioned on latent positions, that is $\tau$, $\bX$ and $\lambda$ in SBM, RDPG and MREG respectively are treated as parameters; furthermore, loops are ignored in MREG. If an adjacency matrix $\bA \sim SBM(\tau,\bB)$ and the block connectivity matrix $\bB$ is positive semidefinite, $\bA$ can also be written as an $RDPG(\bX)$ with $\bX$ having at most $K$ distinct rows. If an adjacency matrix $\bA \sim RDPG(\bX)$, then it is also a $1$-graph $MREG(\lambda_1,h_1,...,h_d)$ with $h_k$ being the normalized $k$th column of $\bX$ and $\lambda_1$ being the vector containing the squared norms of columns of $\bX$. However, a $1$-graph $MREG(\lambda_1,h_1,...,h_d)$ is not necessarily an RDPG graph since $\lambda_1$ could contain negative entries which may result in an indefinite edge probability matrix. \\

The bottom panel of Figure \ref{fig:ven} shows the relationships between the models on multiple graphs. For  RDPG, the graphs are sampled i.i.d. with the same parameters. MREG has the flexibility to have $\lambda$ differ across graphs, which leads to a more generalized model for multiple graphs.  A $d$ dimensional MREG can represent any SBM with $K\leq d$ blocks, in which the block memberships are the same across all the graphs in the population, but possible different connectivity matrices $\bB_i$ for each graph, a common assumption in modeling multilayer and time-varying networks \cite{Han,peixoto2015inferring}. Other models in this setting also allow some vertices of the SBM to change their block memberships over time \cite{Han,ghasemian2016detectability}; in this case, the MREG can still represent those models, but the dimension $d$ may need to increase. Actually, it turns out that if $d$ is allowed to be as large as $\frac{n(n-1)}{2}$, MREG can represent any distribution on binary graphs, which includes distributions in which edges are not independent.   
\begin{theorem}
	Given any distribution $\mathcal{F}$ on graphs and a random adjacency matrix $\bA \sim \mathcal{F}$, there exists a dimension  $d$, a distribution $F$ on $\mathbb{R}^d$, and a set of vectors $\{h_k\}_{k=1}^d$, such that $\bA \sim MREG(F,h_1,...,h_d)$.
	\label{thm:rep}
\end{theorem}
Theorem \ref{thm:rep} implies that MREG is really a semi-parametric model, which can capture any distribution on graphs. One can model any set of graphs by MREG with the guarantee that the true distribution is in the model with $d$ being large enough. However, in practice, a smaller $d$ may lead to better inference performance due to reduction in the dimensionality.

In the next section, we consider the joint embedding algorithm which can be understood as a parameter estimation procedure for MREG.

\section{Joint Embedding}
\subsection{Joint Embedding of Graphs}
The joint embedding method considers a collection of vertex-aligned graphs, and estimates a common embedding space across all graphs and a loading for each graph. Specifically, it simultaneously identifies a subspace spanned by a set of rank one symmetric matrices and projects each adjacency matrix $\bA_i$ into the subspace. The coefficients obtained by projecting $\bA_i$ are denoted by $\hat{\lambda}_i \in \mathbb{R}^d$, which is called the loading for graph $i$. To estimate rank one symmetric matrices and loadings for graphs, the algorithm minimizes the sum of squared Frobenius distances between adjacency matrices and their projections as described below.\\
\begin{definition} Joint Embedding of Graphs (JE). Given $m$ graphs $\{G_i \} _{i=1}^{m}$ with $\bA_i$ being the corresponding adjacency matrix, the $d$-dimensional joint embedding of graphs $\{G_i \} _{i=1}^{m}$ is given by
	\begin{equation}\label{eq:1}
	(\hat{\lambda}_1,...,\hat{\lambda}_m,\hat{h}_1,...,\hat{h}_d) = \underset{\lambda_i,\|h_k\|=1}{\operatorname{argmin}} \sum\limits_{i=1}^{m} \| \bA_i- \sum\limits_{k=1}^{d} \lambda_{i}[k] h_k h_k^T \|  ^2.  
	\end{equation}
	Here, $\| \cdot \|$ denotes the Frobenius norm and $\lambda_{i}[k]$ is the $k$th entry of vector $\lambda_i$.
\end{definition}

To make sure that the model is identifiable and avoid the scaling factor, $h_k$ is required to have norm $1$. In addition, $\{h_k h_k^T\}_{k=1}^d$ must be linearly independent to avoid identifiability issues in estimating $\lambda_i$; however, $\{h_k\}_{k=1}^d$ needs not to be linearly independent or orthogonal. To ease the notations, let us introduce two matrices $\bLambda \in \mathbb{R}^{m \times d}$ and $\bH\in \mathbb{R}^{n \times d}$, where $\lambda_i$ is the $i$th row of $\Lambda$ and $h_k$ is the $k$th row of $\bH$; that is, $\bLambda=[\lambda_1^T,...,\lambda_m^T]$ and $\bH=[h_1,...,h_d]$. The equation \eqref{eq:1} can be rewritten using $\bLambda$ and $\bH$ as
\begin{equation*}
(\hat{\bLambda},\hat{\bH}) = \underset{\bLambda,\|h_k\|=1}{\operatorname{argmin}} \sum\limits_{i=1}^{m} \| \bA_i- \sum\limits_{k=1}^{d} \bLambda_{ik} h_k h_k^T \|  ^2.  
\end{equation*}
Denote the function on the left hand side of the equation by $f(\bLambda,\bH)$ which is explicitly a function of $\lambda_i$s and $h_k$s. There are several alternative ways to formulate the problem. If vector $\lambda_i$ is converted into a diagonal matrix $\bD_i \in \mathbb{R}^{d \times d}$ by putting entries of $\lambda_i$ on the diagonal of $\bD_i$, then solving equation \eqref{eq:1} is equivalent to solving
\begin{equation*}
\begin{aligned}  
& \underset{\bD_i,\|h_k\|=1}{\operatorname{argmin}} 
& & \sum\limits_{i=1}^{m} \| \bA_i- \bH \bD_i \bH^T \|  ^2 \\
& \text{ subject to} 
& &  \bD_i \text{ being diagonal.}
\end{aligned}
\end{equation*}
Equation \eqref{eq:1} can be also viewed as a tensor factorization problem. If $\{\bA_i\}_{i=1}^m$ are stacked in a 3-D array ${\mathbb A} \in \mathbb{R}^{m\times n \times n}$, then solving equation \eqref{eq:1} is also equivalent to
\[  \underset{\bLambda,\|h_k\|=1}{\operatorname{argmin}}  \| {\mathbb A} - \sum\limits_{k=1}^{d} \bLambda_{*k} \otimes h_k \otimes h_k\|  ^2,  \]
where $\otimes$ denotes the tensor product and $\bLambda_{*k}$ is the $k$th column of $\bLambda$. It is well known in the tensor factorization community that the solution to Equation \ref{eq:1} may not necessarily exist for $d \geq 2$. This phenomenon was first found by Bini \textit{et al.} \cite{bini1979n2}, and Silva and Lim gives a characterization all such tensors in the order-$3$ rank-$2$ case \cite{de2008tensor}. Although there may not exist a global minimum, finding the local solution in a compact region still provide significant insights to the data. We design an algorithm which is guaranteed to converge, and provide analysis under the $d=1$ case.  

The joint embedding algorithm assumes the graphs are vertex-aligned and undirected. The vertex-aligned graphs are common in many applications of interest such as neuroimaging \cite{nolte2002human}, multilayer networks \cite{kivela2014multilayer} or time-varying graphs \cite{park2013anomaly}. In case that the graphs are not aligned, graph matching should be performed before the joint embedding \cite{yan2014graduated,park2016encouraging}. The mis-alignments of some vertices will have adverse effects in estimating corresponding latent positions in $\bH$; however, a small number of mis-aligned vertices should not have a big impact in estimating $\bLambda$. If the graphs have weighted edges, the joint embedding can still be applied. Also, the MREG model can be easily extended to weighted graphs by replacing the Bernoulli distribution with other proper distributions. In fact, in the experiment of section 5.3, the graphs are weighted, where the edge weights are the log of fiber counts across regions of brains. In case of directed graphs, to apply the joint embedding, one can symmetrize the graph by removing the direction of edges. Alternatively, $h_k h_k^T$ in equation \eqref{eq:1} can be replaced by $h_k g_k^T$, with  $h_k$ and $g_k$ representing the in and out latent positions respectively. With this modification, equation \eqref{eq:1} becomes the tensor factorization problem \cite{kolda2009tensor}. 

The optimization problem in equation \eqref{eq:1} is similar to Principal Component Analysis (PCA) in the sense of minimizing squared reconstruction error to recover loadings and components \cite{jolliffe2002principal}. However, there are extra symmetries and rank constraints on the components. Specifically, if $h_k h_k^T$ is replaced by a matrix $\bS_k$ in equation \eqref{eq:1}
\begin{equation}
\label{eq:pca}
(\hat{\lambda}_1,...,\hat{\lambda}_m,\hat{\bS}_1,...,\hat{\bS}_d) = \underset{\lambda_i,\bS}{\operatorname{argmin}} \sum\limits_{i=1}^{m} \| \bA_i- \sum\limits_{k=1}^{d} \lambda_{i}[k] \bS_k \|  ^2,  
\end{equation}
the problem can be solved by applying PCA on vectorized adjacency matrices. In this case, there is a $\bS_k$ to estimate for each latent dimension which has $\frac{n(n+1)}{2}$ parameters. Compared to PCA, the joint embedding estimates a rank one matrix $h_k h_k^T$ for each latent dimension which has $n$ parameters, and $h_k$ can be treated as latent positions for vertices, but the joint embedding yields a larger approximation error due to the extra constraints. Similar optimization problems have also been considered in the simultaneous diagonalization literature \cite{flury1986algorithm,ziehe2004fast}. The difference is that the joint embedding is estimating an $n$-by-$d$ matrix $\bH$ by minimizing reconstruction error instead of finding a $n$-by-$n$ non-singular matrix by trying to simultaneously diagonalize all matrices. The problem in equation \eqref{eq:1} has considerably fewer parameters to optimize, which makes it more stable and applicable with $n$ being moderately large. In case of embedding only one graph, the joint embedding is equivalent to the Adjacency Spectral Embedding solved by singular value decomposition \cite{sussman2012consistent}. Next, we describe an algorithm to optimize the objective function $f(\bLambda,\bH)$.  

\subsection{Alternating Descent Algorithm}
The joint embedding of $\{G_i \} _{i=1}^{m}$ is estimated by solving the optimization problem in equation \eqref{eq:1}. There are a few methods proposed to solve similar problems. Alternating least squares (ALS)  is a popular method to solve similar problems \cite{carroll1970analysis,kolda2009tensor}, but often ignore symmetric constraints. Gradient approaches have also been considered for similar problems \cite{tang2009clustering, kolda2015numerical}. We develop an alternating descent algorithm to minimize $f(\bLambda,\bH)$ that combines ideas from both approaches \cite{bezdek2003convergence}. The algorithm can also be understood as a block coordinate descent method with $\bLambda$ and $\bH$ being the two blocks \cite{wright2015coordinate,beck2013convergence}. The algorithm iteratively updates one of $\bLambda$ and $\bH$ while treating the other parameter as fixed. Optimizing $\bLambda$ when fixing $H$ is straightforward, since it is essentially a least squares problem. However, optimizing $\bH$ when fixing $\bLambda$ is hard due to the fact that the problem is non-convex and there is no closed form solution available. In this case, the joint embedding algorithm utilizes gradient information and takes an Armijo backtracking line search strategy to update $\bH$  \cite{nocedal2006numerical}.

Instead of optimizing all columns $\bLambda$ and $\bH$ simultaneously, we consider a greedy algorithm which solves the optimization problem by only considering one column of  $\bH$ at a time. Specifically, the algorithm fixes all estimates for the first $k_0-1$ columns of $\bLambda$ and $\bH$ at iteration $k_0$, and then the objective function is minimized by searching through only the $k_0$th column of $\bLambda$ and $\bH$. That is,
\begin{align}(\hat{\bLambda}_{*k_0},\hat{h}_{k_0}) =  &\underset{\bLambda_{*k_0},\|h_{k_0}\|=1}{\operatorname{argmin}}\nonumber \\
&\sum\limits_{i=1}^{m} \| \bA_i- \sum\limits_{k=1}^{k_0-1} \hat{\bLambda}_{ik} \hat{h}_{k} \hat{h}_{k}^T -\bLambda_{ik_0} h_{k_0} h_{k_0}^T\|  ^2.
\label{eq:2}
\end{align}
When $m=1$, by the Eckart-Young theorem, the solution for $(\hat\bLambda_{1,k_0}, \hat{h}_{k_0})$ is given by the leading eigenvalue and eigenvector of $\bA_1- \sum\limits_{k=1}^{k_0-1} \hat{\bLambda}_{1k} \hat{h}_{k} \hat{h}_{k}^T$ , and hence the solution obtained by our greedy approach coincides with the eigendecomposition of $\bA_1$. If $m>1$ there is no closed-form solution in general, so we develop an optimization method next.

Let $f(\bLambda_{*k_0},h_{k_0})$ denote the sum on the left hand side of the equation. To compute a $d$-dimensional joint embedding $(\hat{\bLambda},\hat{\bH})$, the algorithm iteratively finds one component at a time by solving the optimization for the one dimensional embedding problem \eqref{eq:2}. Once a new component $k_0$ is found, the matrix of loadings $\hat{\bLambda}$ is updated by minimizing the optimization problem \eqref{eq:1} with the first $k_0$ components fixed at $\hat{h}_1,\ldots, \hat{h}_{k_0}$, resulting in the least squares problem
\begin{align}
     (\hat\bLambda_{*1},...,\hat\bLambda_{*k_0})= \underset{\bLambda\in\mathbb{R}^{m\times k_0}}{\operatorname{argmin}} \sum\limits_{i=1}^{m} \| \bA_i- \sum\limits_{k=1}^{k_0} {\bLambda}_{ik} \hat{h}_{k} \hat{h}_{k}^T\|  ^2.
\label{eq:optimization-globalsolution}
\end{align}
This step finds the projection of the graphs into the space spanned by the first $k_0$ components $\{\hat{h}_k\hat{h}_k^T\}_{k=1}^{k_0}$. The algorithm solves the $d$-dimensional embedding problem by iteratively updating $\hat{\bH}$ and $\hat{\bLambda}$ using equations \eqref{eq:2} and \eqref{eq:optimization-globalsolution}.

There are a few advantages in iteratively solving one dimensional embedding problems. First, there are fewer parameters to fit at each iteration, since the algorithm is only allowed to vary $h_{k_0}$ at iteration $k_0$. This makes initialization and optimization steps much easier compared to optimizing all columns of $\bH$ simultaneously. Second, it implicitly enforces an ordering on the columns of $\bH$. This ordering allows us to select the top few columns of $\bLambda$ and $\bH$ in cases where model selection is needed after the joint embedding. Third, it allows incremental computation. If $d$ and $d'$ dimensional joint embeddings are both computed, the first $\min(d,d')$ columns of $\hat{\bH}$ will be the same. Fourth, although the global rank-$d$ tensor approximation problem does not always have a solution for $d\geq2$, a one dimensional embedding that corresponds to a rank-1 tensor approximation always have a solution \cite{de2008tensor}. Finally, based on numerical experiments, the difference between optimizing iteratively and optimizing all the parameters when $d$ is small is negligible; however, the iterative algorithm yields a slightly smaller objective function when $d$ is large. The disadvantage of optimizing each column separately is that the algorithm is more likely to end up at a local minimum when the objective function is structured not in favor of embedding iteratively. In practice, this problem can be mitigated by running the joint embedding algorithm several times with random initializations. 

To find $\bLambda_{*k_0}$ and $h_{k_0}$ in equation \eqref{eq:2}, the algorithm needs to evaluate two derivatives: $\frac{\partial f}{\partial h_{k_0}}$ and $\frac{\partial f}{\partial \bLambda_{i k_0}}$. Denote by $\bR_{ik_0}$ the residual matrix after iteration $k_0-1$ which is $\bA_i- \sum\limits_{k=1}^{k_0-1}\hat{\bLambda}_{ik} \hat{h}_{k} \hat{h}_{k}^T$. The gradient of the objective function with respect to $h_{k_0}$ is given by
\begin{equation} \label{eq:3}
\frac{\partial f}{\partial h_{k_0}} = -4\sum\limits_{i=1}^{m}  \bLambda_{ik_0} (\bR_{ik}-\bLambda_{ik_0} h_{k_0} h_{k_0}^T)  h_{k_0}.
\end{equation}
The derivative of the objective function with respect to $\bLambda_{i k_0}$ is given by
\[\frac{\partial f}{\partial \bLambda_{i k_0}}= -2 \langle \bR_{ik}-\bLambda_{ik_0} h_{k_0} h_{k_0}^T,h_{k_0} h_{k_0}^T\rangle.\]
Setting the derivative to $0$ yields
\begin{equation}  \label{eq:4}
\hat{\bLambda}_{i k_0} = \langle \bR_{ik}, h_{k_0} h_{k_0}^T \rangle,
\end{equation}
where $\langle \cdot , \cdot \rangle$ denotes the inner product. 

Once a new component $\hat{h}_{k_0}\hat{h}_{k_0}^T$ is obtained, the algorithm proceeds to update $\bLambda$ by solving the optimization problem \eqref{eq:optimization-globalsolution}. Note that the problem can be split into $m$ least square subproblems, one for each graph $\bA_i$, of the form
 \begin{align*}
     (\hat\bLambda_{i1},...,\hat\bLambda_{ik_0}) & = \underset{\bLambda\in\mathbb{R}^{k_0}}{\operatorname{argmin}}  \| \bA_i- \sum\limits_{k=1}^{k_0} {\bLambda}_{k} \hat{h}_{k} \hat{h}_{k}^T\|  ^2,
\end{align*} 
for each $i=1,\ldots,m$. By obtaining the derivative with respect to $\bLambda$ for all $i=1,\ldots,m$ and setting them equal to zero,  $\bLambda$ is updated by solving $m$ systems of  linear equations with $k_0$ variables each. Let $\bGamma$ be a $k_0\times k_0$ matrix and $\bPsi$ be a ${k_0\times m}$ matrix such that $\bGamma_{kj} = (\hat{h}_{j}^T\hat{h}_{k})^2$ and $\bPsi_{ki} = (\hat{h}_{k}^T\bA_i\hat{h}_{k})$. The algorithm updates the first $k_0$ columns of $\bLambda$ by solving
\begin{align}
\bGamma \hat\bLambda_{*,1:k_0}^T = \bPsi. \label{eq:update-lambda}
\end{align}

The joint embedding algorithm alternates between updating $\hat{\bLambda}_{* k_0}$ and $\hat {h}_{k_0}$ according to equation \eqref{eq:3} and \eqref{eq:4}, after which the values of all the loadings $(\hat{\bLambda}_{* 1},\ldots,\hat{\bLambda}_{* k_0})$ are updated using equation \eqref{eq:update-lambda}.  Algorithm \ref{alg:je} describes the general procedure to compute the $d$-dimensional joint embedding of graphs $\{G_i\}_{i=1}^m$. The algorithm outputs two matrices: $\hat{\bLambda}$ and $\hat{\bH}$. The rows of $\hat{\bLambda}$ denoted by $\{\hat{\lambda}_i\}_{i=1}^m$ can be treated as estimates of $\{\lambda_i\}_{i=1}^m$ in MREG and features for graphs. Columns of $\hat{\bH}$ denoted by $\{\hat{h}_k\}_{k=1}^d$ are estimates of $\{h_k\}_{k=1}^d$. If a new graph $G$ is observed with adjacency matrix $\bA$, $\bA$ can be projected into the linear space spanned by $\{\hat{h}_k \hat{h}_k^T\}_{k=1}^{d}$ to obtain features for the graph. 

In case of $\bA_i$ being large, the updating equations \eqref{eq:3} and \eqref{eq:4} are not practical due to $h_k h_k^T$ and $\bR_{ik}$ being large and dense. However, they can be rearranged to avoid explicit computation of $h_k h_k^T$ and $\bR_{ik}$. Equation \eqref{eq:3} becomes
\begin{eqnarray*}
\frac{\partial f}{\partial h_{k_0}} 
&=&-4\sum\limits_{i=1}^{m}  \bLambda_{ik_0} \bA_i h_{k_0} + 4 \sum\limits_{i=1}^{m}  \bLambda_{ik_0} \sum\limits_{k=1}^{k_0-1}  \bLambda_{ik} (h_k^T h_{k_0}) h_k \\
& &+ 4\sum\limits_{i=1}^{m}  \bLambda_{ik_0}^2  h_{k_0}.
\end{eqnarray*}
Similarly,  equation \eqref{eq:4} can be rewritten as
\begin{align*}  
\hat{\bLambda}_{i k_0} 
&= h_{k_0} ^T \bA_i h_{k_0} - \sum\limits_{k=1}^{k_0-1} \bLambda_{ik} (h_{k_0} ^T h_k)^2.
\end{align*}
Based on the rearranged equations, efficiently evaluating matrix vector product $\bA_i h_{k_0}$ is needed to calculate the derivatives. This can be completed for a variety of matrices, in particular, sparse matrices \cite{bell2009implementing}.  

The Algorithm \ref{alg:je} is guaranteed to converge to a stationary point. Specifically, at the termination of iteration $k_0$, $\frac{\partial f}{\partial h_{k_0}} \approx 0$ and $\frac{\partial f}{\partial \bLambda_{i k_0}} \approx 0$. First, $\frac{\partial f}{\partial \bLambda_{i k_0}} \approx 0$ is ensured due to exact updating by equation \eqref{eq:4}. Second notice that updating according to equation \eqref{eq:3} and \eqref{eq:4} always decreases the objective function. Due to the fact that $h_{k_0}$ lies on the unit sphere and the objective is twice continuous differentiable, $\frac{\partial f}{\partial h_{k_0}}$ is Lipschitz continuous. This along with Armijo backtracking line search guarantees a "sufficient" decrease $c\|\frac{\partial f}{\partial h_{k_0}}\|^2$ in objective function each time when the algorithm updates $h_{k_0}$ \cite{nocedal2006numerical}, where $c$ is a constant independent of $h_{k_0}$. Since the objective function is bounded below by $0$, this implies convergence of gradient, that is $\frac{\partial f}{\partial h_{k_0}} \rightarrow 0$.

The joint embedding of graphs model requires for identifiability that 
the basis of the embedding space $\{h_kh_k^T\}_{k=1}^m$ is linearly independent. This condition also ensures that equation \eqref{eq:optimization-globalsolution} has a unique solution. Although the optimization problem \eqref{eq:2} does not directly enforce this constraint, the next theorem guarantees that the components obtained by our method are linearly independent. The proof is given in the appendix.

\begin{theorem}
Suppose that  $\hat{\bLambda}=(\hat\bLambda_{*1},...,\hat\bLambda_{*k_0})$ and $\hat{\bH}=[\hat h_1,...,\hat h_{k_0}]$ are  obtained by iteratively solving \eqref{eq:2} and \eqref{eq:optimization-globalsolution}
for $k_0$ components, and the first $k_0-1$ components $\{\hat{h}_k \hat{h}_k^T\}_{k=1}^{k_0-1}$ are linearly independent. Then $\{\hat{h}_k \hat{h}_k^T\}_{k=1}^{k_0}$ are also linearly independent, or $\hat{\bLambda}_{ik_0}=0$ for all $i=1,\ldots,m$. 
\label{theo:optimization}
\end{theorem}

In general, the objective function may have multiple stationary points due to non-convexity. Therefore, the joint embedding algorithm is sensitive to initializations. Actually, like many of the problems in tensor factorization, finding the global minimum in joint embedding is NP-Hard \cite{hillar2013most}.  When time permits, we recommend running the joint embedding several times including many random initializations. In Section 5.1, we study the effects of different initialization approaches through a numerical simulation experiment. For other simulation and real experiments, we initialize $\hat {h}_{k_0}$ using the top left singular vector of the average residual matrix $\sum \bR_{ik_0}/m$. Computing this vector is computationally cheap compared to the computational cost of the joint embedding when the number of graphs $m$ is larger than 1. When $m=1$, the SVD initialization and the joint embedding solution coincide, and when $m>1$ the SVD initialization gives a good approximation to the solution if all the graphs are identically distributed. The optimization algorithm described above may not be the fastest approach to solving the problem; however, numerical optimization is not the focus of this paper. Based on results from numerical applications, our approach works well in estimating parameters and extracting features for subsequent statistical inference. Next, we discuss some variations of the joint embedding algorithm.
\begin{algorithm}
	\caption{Joint Embedding}
	\label{alg:je}
	\begin{algorithmic}[1]
		\Procedure{Find joint embedding $\hat{\bLambda},\hat{\bH}$ of $\{\bA_i\}_{i=1}^m$}{}
		\State Set residuals: $\bR_{i1}=\bA_i$
		\For{$k=1:d$ }
		\State Initialize $h_k$ and $\bLambda_{*k}$ 
		\While{not convergent}
		\State Fixing $\bLambda_{*k}$, update $h_k$ by gradient descent \eqref{eq:3}
		\State Project $h_k$ back to the unit sphere
		\State Fixing $h_k$, update $\bLambda_{*k}$ by \eqref{eq:4}
		\State Compute objective $\sum\limits_{i=1}^{m} \| \bR_{ik}-  \bLambda_{ik} h_k h_k^T \|^2$
		\EndWhile
        \State Update $\bLambda$ by solving \eqref{eq:update-lambda}
		\State Update residuals: $\bR_{i(k+1)}=\bA_{i}-\sum_{j=1}^k \bLambda_{ij} h_jh_j^T$
		\EndFor
		\State Output $\hat{\bLambda}=[\bLambda_{*1},...,\bLambda_{*d}]$ and $\hat{\bH}=[h_1,...,h_d]$
		\EndProcedure
	\end{algorithmic}
\end{algorithm}

\subsection{Variations}
The joint embedding algorithm described in the previous section can be modified to accommodate several different settings. 

\textbf{Variation 1.} When all graphs come from the same distribution, we can force estimated loadings $\hat{\lambda}_i$ to be equal across all graphs. This is useful when the primary inference task is to extract features for vertices. Since all graphs share the same loadings, with slightly abusing notations, let $\bLambda$ be a vector in $\mathbb{R}^d$ and the optimization problem becomes
\[ (\hat{\bLambda},\hat{\bH}) = \underset{\bLambda,\|h_k\|=1}{\operatorname{argmin}} \sum\limits_{i=1}^{m} \| \bA_i- \sum\limits_{k=1}^{d} \bLambda_{k} h_k h_k^T \|  ^2,  \] 
which is equivalent to 
\[ (\hat{\bLambda},\hat{\bH}) = \underset{\bLambda,\|h_k\|=1}{\operatorname{argmin}} \| \frac{1}{m}\sum\limits_{i=1}^{m}\bA_i - \sum\limits_{k=1}^{d} \bLambda_{k} h_k h_k^T \|  ^2.  \] 
Therefore, the optimization problem can be solved exactly by finding the singular value decomposition of the average adjacency matrix $\frac{1}{m}\sum\limits_{i=1}^{m}\bA_i$. \\
\textbf{Variation 2.} When there is a discrete label $y_i \in \mathbb{Y}$ associated with the graph $G_i$ available, we may require all loadings $\hat{\lambda}_i$ to be equal within class. Let $\bLambda \in \mathbb{R}^{|\mathbb{Y}| \times d}$, the optimization problem becomes
\[ (\hat{\bLambda},\hat{\bH}) = \underset{\bLambda,\|h_k\|=1}{\operatorname{argmin}} \sum\limits_{i=1}^{m} \| \bA_i- \sum\limits_{k=1}^{d} \bLambda_{y_i k} h_k h_k ^T \|  ^2.  \] 
In this case, when updating $\bLambda$ as in equation \eqref{eq:4}, the algorithm should average $\bLambda_{y k}$ within the same class, that is
\[
\hat{\bLambda}_{y k} = \left( \sum\limits_{i=1}^{m} \mathbb{I}\{y_i = y\} \langle \bR_{i k}, h_{k_0} h_{k_0}^T \rangle\right)/\left(\sum\limits_{i=1}^{m} \mathbb{I}\{y_i = y\}\right).\] \\
\textbf{Variation 3.} In some applications, we may require all $\bLambda_{ik}$ to be greater than $0$, as in non-negative matrix factorization. One advantage of this constraint is that graph $G_i$ may be automatically clustered  based on the largest entry of $\hat{\lambda}_{i}$. In this case, the optimization problem is
\[ (\hat{\bLambda},\hat{\bH}) = \underset{\bLambda \geq 0,\|h_k\|=1}{\operatorname{argmin}} \sum\limits_{i=1}^{m} \| \bA_i- \sum\limits_{k=1}^{d} \bLambda_{ik} h_k h_k ^T \|  ^2.  \] 
To guarantee nonnegativity, the algorithm should use nonnegative least squares in updating $\bLambda$ \cite{kim2008nonnegative}. Furthermore, a constraint on the number of non-zero elements in $i$th row of $\bLambda$ can be added as in K-SVD \cite{aharon2006rm}, and a basis pursuit algorithm could be used to update $\bLambda$ \cite{chen2001atomic, tropp2007signal}. Next, we discuss some theoretical properties of the MREG model and joint embedding when treated as a parameter estimation procedure for the model.

\section{Theoretical Results}
In this section, we consider a simple setting where graphs follow a $1$-dimensional MREG model, that is $\{(\lambda_i,\bA_i)\} _{i=1}^m \sim MREG(F,h_1)$. The $1$-dimensional joint embedding is well defined in this case, that is $\hat{\lambda}_i$ and $\hat{h}_1$ defined in Equation \ref{eq:1} is guaranteed to exist. Under this MREG model, the joint embedding of graphs can be understood as estimators for parameters of the model. Specifically, $\hat{\lambda}_i$ and $\hat{h}_1$ are estimates of $\lambda_i$ and $h$. We prove two theorems concerning the asymptotic behavior of estimator $\hat{h}_1$ produced by joint embedding. 

Let $\hat{h}_1^m$ denote the estimates based on $m$ graphs and define functions $\rho$, $D_m$ and $D$ as below: 
\[ \rho(\bA_i,h)= \|\bA_i- \langle \bA_i,h h^T \rangle h h^T\|^2, \]
\[ D_m(h,h_1) =\frac{1}{m}\sum_{i=1}^{m} \rho(\bA_i,h), \]
\[ D(h,h_1) = \EE(\rho(\bA_i,h)). \]
One can understand $D_m$ and $D$ as sample and population  approximation errors respectively. By equation \eqref{eq:1},
\begin{align*} 
\hat{h}_1^m = \underset{\|h\| =1}{\operatorname{argmin}} \text{ }   \underset{\lambda_i}{\operatorname{argmin}} \sum_{i=1}^{m} \|\bA_i - \lambda_i h h^T\|.
\end{align*}
By equation \eqref{eq:4}, 
\[\langle \bA_i,hh^T \rangle=\underset{\lambda_i}{\operatorname{argmin}} \sum_{i=1}^{m} \|\bA_i - \lambda_i h h^T\|.\]
Therefore,
\[\hat{h}_1^m = \underset{\|h\| =1}{\operatorname{argmin}} \text{ } D_m(h,h_1). \]
The first theorem states that $\hat{h}_1^m$  converges almost surely to a global minimum of $D(h,h_1)$. Alternatively, the theorem implies the sample minimizer converges to the population minimizer.
\begin{theorem}
	\label{thm:je1}
The estimator $\hat{h}_1^m$ converges almost surely to the set of global minimizers of $D(h,h_1)$ as $m$ goes to infinity. That is, 
	\[ \mathbb{P}\left(\lim_{m\rightarrow\infty}\hat{h}_1^m\in\underset{h}{\operatorname{argmin}} D(h,h_1)\right) =1. \]
\end{theorem}

Theorem \ref{thm:je1} ensures that, in the limit, $\hat{h}^m$ is arbitrarily close to parameters in the set of global minimizer of $D(h,h_1)$. However, the global minimizer is definitely not unique due to the symmetry up to sign flip of $h$, that is $D(h,h_1)=D(-h,h_1)$ for any $h$. This problem can be addressed by forcing an orientation of $\hat{h}_1^m$ or stating that the convergence is up to a sign flip. In this case, Theorem \ref{thm:je1} does not apply. We are currently only certain that when all graphs are from the Erdos-Renyi random graph model, the global minimizer is unique up to a sign flip. The next theorem concerns the asymptotic bias of $h'$. It gives a bound on the difference between the population minimizer  $h'$ and the truth $h_1$.
\begin{theorem}
	\label{thm:je2}
	If $h'$ is a minimizer of $D(h,h_1)$, then 
	\[\|h'-h_1\| \leq \frac{2 \EE(\lambda_i)}{\EE(\lambda_i^2)(h_1^T h')^2}. \]
\end{theorem}

To see an application of Theorem \ref{thm:je2}, let us consider the case in which all graphs are Erdos-Renyi graphs with $100$ vertices and edge probability of $0.5$. Under this setting, Theorem \ref{thm:je2} implies  $\|h'-h_1\| \in [0,0.04] \cup [1.28,1.52]$. The second interval is disturbing. It is due to the fact that when $h_1^T h'$ is small, the bound is useless. We provide some insights as to why the second interval is there and how we can get rid of it with additional assumptions. In the proof of Theorem \ref{thm:je2}, we show that the global optimizer $h'$ satisfies
\[h'= \underset{\|h\| =1}{\operatorname{argmax}} \text{ } \EE(\langle \bA_i,h h^T \rangle ^2). \]
Taking a closer look at $E(\langle \bA_i,h h^T \rangle ^2)$,
\begin{align*}
\EE(\langle \bA_i,h h^T \rangle ^2) &= \EE(\langle \bP_i,h h^T \rangle ^2)+\EE(\langle \bA_i-\bP_i,h h^T \rangle ^2) \\
&=\EE(\lambda_i^2)(h_1^T h)^4+\EE((h^T (\bA_i-\bP_i)h) ^2).
\end{align*}
Therefore, 
\[h'= \underset{\|h\| =1}{\operatorname{argmax}} \text{ } \EE(\lambda_i^2)(h_1^T h)^4+\EE((h^T (\bA_i-\bP_i)h) ^2) .\]
We can see that $\EE(\lambda_i^2)(h_1^T h)^4$ is maximized when $h=h_1$; however, the noise term $\EE((h^T (\bA_i-\bP_i)h) ^2)$ is generally not maximized at $h=h_1$. If $n$ is large, we can apply a concentration inequality to $(h^T (\bA_i-\bP_i)h) ^2$ and have an upper bound on $\EE((h^T (\bA_i-\bP_i)h) ^2)$. If we further assume $A_i$ is not too sparse, that is $\EE(\lambda_i^2)$ grows with $n$ fast enough, then the sum of these two terms is dominated by the first term. This provides a way to have a lower bound on $h_1^T h'$. We may then replace the denominator of the bound in Theorem \ref{thm:je2} by the lower bound. In general, if $n$ is small, the noise term may cause $h'$ to differ from $h_1$ by a significant amount. In this chapter, we focus on the case that $n$ is fixed. The case that $n$ goes to infinity for Random Dot Product Graph is considered in \cite{athreya2013limit}.

The two theorems above concern only the estimation of $h_1$, but not $\lambda_i$. Based on equation \eqref{eq:4}, the joint embedding estimates $\lambda_i$ by
\[\hat{\lambda}_i^m= \langle \bA_i,\hat{h}_1^m \hat{h}_1^{m T} \rangle. \]
When $m$ goes to infinity, we can apply Theorem \ref{thm:je1},
\[\hat{\lambda}_i^m = \langle \bA_i,\hat{h}_1^m \hat{h}_1^{mT} \rangle \overset{a.s.}{\rightarrow} \langle \bA_i,h' h'^T \rangle = h'^T \bA_i h'.\]
Then, applying the bound on $\|h'-h_1\|$ derived in Theorem \ref{thm:je2} and utilizing the fact that $h^T \bA_i h$ is continuous in $h$, we can obtain an upper bound on $|\hat{\lambda}_i^m - h_1^T \bA_i h_1|$. When $\bA_i$ is large, $h_1^T \bA_i h_1$ is concentrated around $\lambda_i$ with high probability. As a consequence, with high probability $|\hat{\lambda}_i^m - \lambda_1|$ is small. In the next section, we demonstrate properties and utilities of the joint embedding algorithm through experiments. 

\section{Experiments}
Before going into details of our experiments, we want to discuss how to select the dimensionality $d$ of the joint embedding. Estimating $d$ is an important model selection question that has been studied for years under various settings \cite{kohavi1995study}. Model selection is not the focus of this paper, but we still face this problem in numerical experiments. In the simulation experiments of this section, we assume $d$ is known to us and simply set the dimensionality estimate $\hat{d}$ equal to $d$. In the real data experiment, we recommend two approaches to determine $\hat{d}$. Both approaches require first running the $d'$-dimensional joint embedding algorithm, where $d'$ is sufficiently large. We then plot the objective function versus dimension, and determine $\hat{d}$ to be where the objective starts to flatten out. Alternatively, we can plot $\{\hat{\bLambda}_{ik}\}_{i=1}^m$ for $k=1,...,d'$, and select $\hat{d}$ when the loadings start to look like noise with $0$ mean. These two approaches should yield a similar dimensionality estimate of $\hat{d}$. 

\subsection{Simulation Experiment 1: Joint Embedding Under a Simple Model}
\begin{figure}[!htb]
	\centering
	\includegraphics[scale=0.6,width=3.0in]{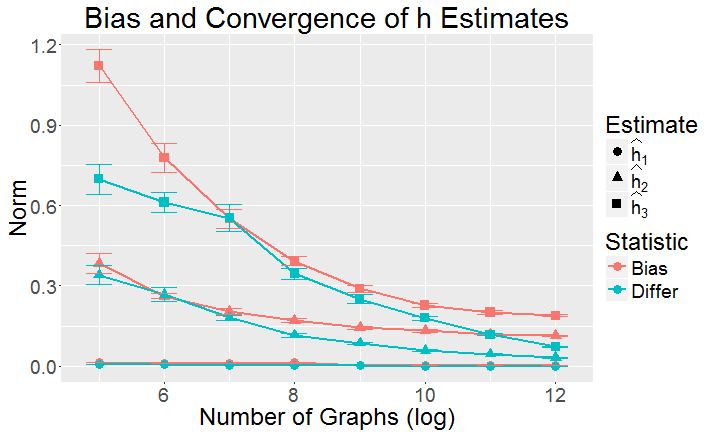}
	\caption{Mean bias ($\|\hat{h}^m_k-h_k\|$) and mean difference between estimates ($\|\hat{h}_k^m-\hat{h}_k^{m/2}\|$) across $20$ simulations are shown. The standard errors are also given by error bars. The graphs are generated from a $3$-dimensional MREG model as described in section 5.1. $\hat{h}_k^m$ has small asymptotic bias; however, it seems to converge as $m$ increases.}
	\label{fig:db}
\end{figure}

In the first experiment, we present a simple numerical example to demonstrate some properties
of the joint embedding procedure as the number of graphs grows. We repeatedly generate graphs with $20$ vertices from $3$-dimensional MREG, where  $\lambda_i[1] \sim \text{Uniform}(8,16)$, $\lambda_i[2] \sim \text{Uniform}(0,2)$ and $\lambda_i[3] \sim \text{Uniform}(0,1)$,  with
\begin{align*}
h_1 &=[1,1,1,...,1]/\sqrt{20}  \\
h_2&=[1,-1,1,-1,1,-1,...,-1] /\sqrt{20}  \\
h_3&=[1,1,-1,-1,1,1,-1,-1,...,-1]/\sqrt{20}. 
\end{align*}
We keep doubling the number of graphs $m$ from $2^4$ to $2^{12}$. At each value of $m$, we compute the $3$-dimensional joint embedding of graphs. Let the estimated parameters based on $m$ graphs be denoted by $\hat{\lambda}_i^m$ and $\hat{h}_k^m$. Two quantities based on $\hat{h}_k^m$ are calculated. The first is the norm difference between the current $h_k$ estimates and the previous estimates, namely $\|\hat{h}_k^m-\hat{h}_k^{m/2}\|$. This provides numerical evidence for the convergence of our principled estimation procedure. The second quantity is $\|\hat{h}^m_k-h_k\|$. This investigates whether $\hat{h}_k$ is an unbiased estimator for $h_k$. The procedure described above is repeated $20$ times. Figure \ref{fig:db} presents the result. \\

Based on the plot, the norm of differences $\|\hat{h}^m_k-\hat{h}_k^{m/2}\|$ seem to converge to $0$ as $m$ increases. This suggests the convergence of $\hat{h}_1^m$. Second, we notice that the bias $\|\hat{h}^m_2-h_2\|$ and $\|\hat{h}^m_3-h_3\|$ do not converge to $0$; instead, it stops decreasing at around $0.1$ and $0.2$ respectively. This suggests that $\hat{h}_k$ is an asymptotically biased estimator for $h_k$. Actually, this is as to be expected: when there are infinitely many nuisance parameters present, Neyman and Scott demonstrate that maximum likelihood estimator is inconsistent \cite{neyman1948consistent}. In our case, there are infinitely many $\lambda_i$ as $m$ grows; therefore, we do not expect the joint embedding to provide an asymptotic consistent estimate of $h_k$. \\

\begin{figure}[!htb]
	\centering
	\includegraphics[scale=0.6,width=3.0in]{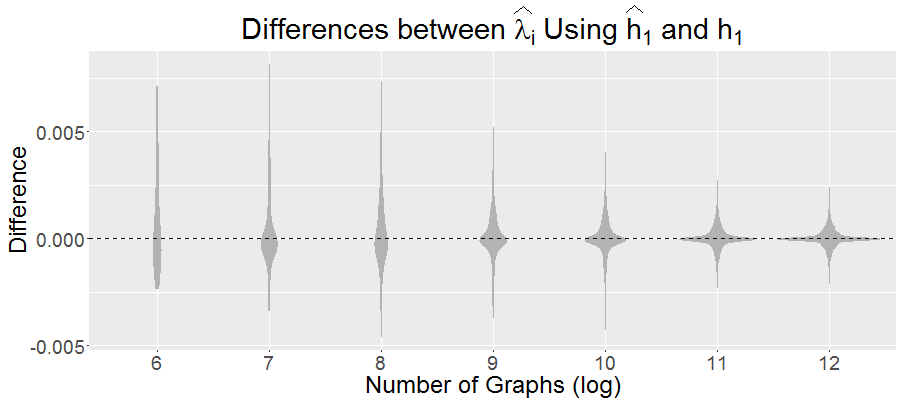}
	\caption{Distribution of differences between $\hat{\lambda}_i[1]$ estimated using $\hat{h}_1^m$ and $h_1$. The graphs are generated from the $3$-dimensional MREG model as described in section 5.1. The differences are small due to the fact that $\hat{h}_1^m$ and $h_1$ are close.}
	\label{fig:ld}
\end{figure}

In applications such as clustering or classifying multiple graphs, we may be not interested in $\hat{h}_k$. $\hat{\lambda}_i$ is of primary interest, which provides information specifically about the graphs $G_i$. Here, we consider two approaches to estimate $\lambda_i[1]$. The first approach is estimating $\lambda_i[1]$ through joint embedding, that is
\[ \hat{\lambda}_i[1] = \langle \bA_i,  \hat{h}^m_1 \hat{h}^{m T}_1 \rangle. \]
The second approach estimates $\lambda_i$ by assuming $h_1$ is known. In this case, equation \eqref{eq:4} gives 
\[ \hat{\lambda}_i[1] = \langle \bA_i,  h_1 h_1^T \rangle. \]
$\hat{\lambda}_i[1]$ calculated this way can be thought as the 'oracle' estimate. Figure \ref{fig:ld} shows the differences in estimates provided by two approaches. Not surprisingly, the differences are small due to the fact that $\hat{h}_1^m$ and $h_1$ are close. \\

Next, we investigate the effects of four different initialization approaches. The first approach utilizes SVD on average residual matrix to initialize $h_k$ at each iteration. The second approach  randomly samples independent Gaussian variable for each entry of $h_k$. The third approach takes the best initialization among $10$ random initializations. The fourth approach initializes $h_k$ using the truth. To compare these approaches, we generate $16$ graphs from the MREG model and jointly embed them with four different initializations. Then, another $16$ graphs are generated and the objective function on these graphs are evaluated using $\hat{\bH}$ estimated by joint embedding. This procedure is repeated $100$ times. Mean objective function and total running time with standard error of these four approaches are shown in Table \ref{tb:inn}.
\begin{table}
	\centering
	\begin{tabular}{ |l|l|l| } 
		\hline
		Initialization & Objective & Running time (sec.) \\ \hline
		SVD & $375.22(1.21)$ & $8.3(1.0)$ \\ \hline
		$1$ Random  & $383.29(1.60)$ & $9.2(1.4)$ \\ \hline
		Best of $10$ Random  & $379.63(1.39)$ & $96.5(5.3)$  \\ \hline
		Truth  & $374.69(1.22)$ & $7.8(1.0)$ \\
		\hline
	\end{tabular}
	\caption{Objective function and running time of four initialization approaches. The standard error is shown in parenthesis. SVD and truth initializations are significantly better than random initializations on this scenario.}
	\label{tb:inn}
\end{table}
Based on Wilcoxon signed-rank test, SVD and truth initializations are significantly better than random initializations on this scenario. For the rest experiments, the initialization is completed by SVD. 

\subsection{Simulation Experiment 2: Joint Embedding to Classify Graphs}
In this experiment, we consider the inference task of classifying graphs.  We have $m$ pairs $\{(\bA_i,y_i)\}_{i=1}^{m}$ of observations. Each pair consists of an adjacency matrix $\bA_i \in \{0,1\}^{n \times n}$ and a label $y_i \in [K]$. Furthermore, all pairs are assumed to be independent and identically distributed according to an unknown distribution $\mathbb{F}_{\bA,y}$, that is
\[(\bA_1,y_1),(\bA_2,y_2),...,(\bA_m,y_m) \overset{i.i.d.}{\sim} \mathbb{F}_{A,y}. \] 
The goal is to find a classifier $g$ which is a function $g:\{0,1\}^{n \times n} \rightarrow [K]$ that has a small classification error $L_g=P(g(\bA)\neq y)$. \\ 

\noindent We consider a binary classification problem where $y$ takes value $1$ or $2$. $200$ graphs with $100$ vertices are independently generated. The graphs are sampled from a $2$-dimensional MREG model. Let $h_1$ and $h_2$ be two vectors in $\mathbb{R}^{100}$, and \[h_1=[0.1,...,0.1]^T \text{ , and } h_2=[-0.1,...,-0.1,0.1,...,0.1]^T. \] 
Here, $h_2$ has $-0.1$ as its first $50$ entries and $0.1$ as its last $50$ entries. Graphs are generated according to the MREG model, 
\begin{equation}
\{(\lambda_i,\bA_i)\}_{i=1}^{200} \sim MREG(F,h_1,h_2),
\label{eq:simu}
\end{equation}
where $F$ is a mixture of two point masses with equal probability, 
\[ F = \frac{1}{2}\mathbb{I} \{\lambda=[25,5]\} + \frac{1}{2}\mathbb{I} \{\lambda=[22.5,2.5]\}.\]
We let the class label $y_i$ indicate which point mass $\lambda_i$ is sampled from. In terms of SBM, this graph generation scheme is equivalent to 
\[ A_i|y_i=1 \sim  SBM((1,...,1,2,...,2),\begin{bmatrix} 0.3 & 0.2 \\ 0.2 & 0.3 \\ \end{bmatrix})  \]
\[ A_i|y_i=2 \sim  SBM((1,...,1,2,...,2),\begin{bmatrix} 0.25 & 0.2 \\ 0.2 & 0.25 \\ \end{bmatrix}).\]

\begin{figure}[!htb]
	\centering
	\includegraphics[scale=0.6,width=3.0in]{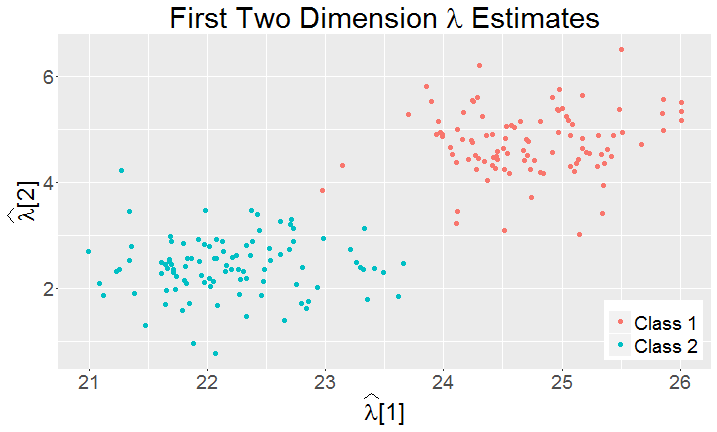}
	\caption{Scatter plot of loadings computed by jointly embedding $200$ graphs. The graphs are generated from the $2$-dimensional MREG model as described in equation \eqref{eq:simu}. The loadings of two classes are separated after being jointly embedded. }
	\label{fig:load}
\end{figure}

\noindent To classify graphs, we first jointly embed $200$ graphs. The first two dimensional loadings are shown in Figure \ref{fig:load}. We can see two classes are separated after being jointly embedded. 
Then, a $1$-nearest neighbor classifier (1-NN) is constructed based on loadings $\{\hat{\lambda}_i\}_{i=1}^m$.\\

\begin{figure}[!htb]
	\centering
	\includegraphics[scale=0.6,width=3.0in]{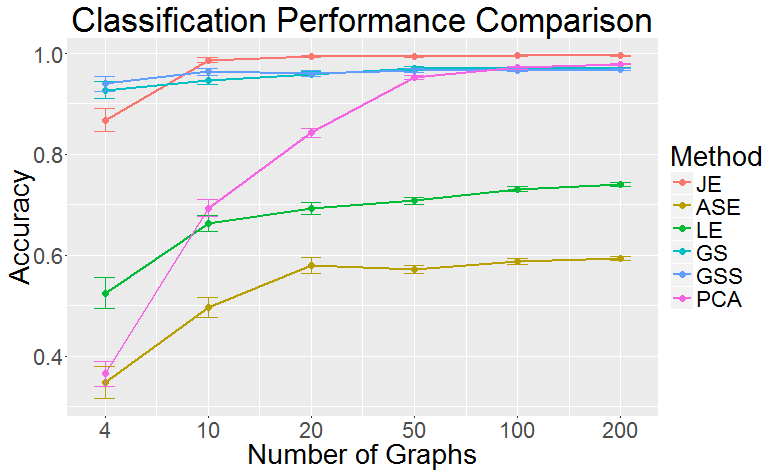}
	\caption{Mean classification accuracy of joint embedding, Adjacency Spectral Embedding, Laplacian Eigenmap, Graph Statistics, Graph Spectral Statistics, and PCA with their standard errors are shown. The graphs are generated from a $2$-dimensional MREG model as described in the equation \eqref{eq:simu}. The features are first extracted using methods described above; subsequently, we apply a $1$-NN to classify graphs. For each value of $m$, the simulation is repeated $100$ times. ASE, LE, GS and GSS do not take advantage of increasing sample size in the feature extraction step. PCA has poor performance when the sample size is small. Joint embedding takes advantage of increasing sample size and outperforms other approaches when given more than $10$ graphs. }
	\label{fig:acc}
\end{figure}

We compare classification performances of using the joint embedding to extract features to five other feature extraction approaches: Adjacency Spectral Embedding, Laplacian Eigenmap, Graph Statistics, Graph Spectral Statistics, and PCA. For Adjacency Spectral Embedding (ASE) and Laplacian Eigenmap (LE), we first embed each adjacency matrix or normalized Laplacian matrix and then compute the Procrustes distance between embeddings. For Graph Statistics (GS), we compute topological statistics of graphs considered by Park \textit{et al.} in \cite{park2013anomaly}. For Graph Spectral Statistics (GSS), we compute the eigenvalues of adjacency matrices and treat them as features \cite{dorogovtsev2003spectra}. For PCA, we vectorize the adjacency matrices and compute the factors through SVD. After the feature extraction step, we also apply a $1$-NN rule to classify graphs. We let the number of graphs $m$ increase from $4$ to $200$. For each value of $m$, we repeat the simulation $100$ times. Figure \ref{fig:acc} shows the result. The joint embedding takes advantage of increasing sample size and outperforms other approaches when given more than $10$ graphs. 

\subsection{Real Data Experiment 1: Subject classification on HNU1 data}
In this section, we use the HNU1 data \cite{zuo2014open} to classify graphs from different subjects. These data consist of diffusion tensor imaging (DTI) records of the brain of 30 healthy different subjects, each of which was scanned 10 times over a period of one month. Each scan was processed with the NeuroData's MRI to Graphs (NDMG) pipeline \cite{kiar2016ndmgcode} using the Talairach atlas \cite{lancaster1997talairach} to register the vertices,  and  we  obtained a sample  of 300 graphs (10 graphs per subject), each of which is composed of 1105 vertices.

It has been suggested that the information encoded in patterns of brain connectivity can uniquely identify different subjects \cite{discrim, finn2015functional}, and there is some evidence of low-rank structure in those differences \cite{Tang2016-rc,sripada2018fundamental}. We study how low-dimensional representations can capture inter-individual variability by using the HNU1 data to classify subject scans. The joint embedding of graphs is an ideal method for this task, since it provides a low-rank representation of each of the graphs which is jointly learned from the sample.

The information encoded in the adjacency matrices can be used to accurately discriminate between different individuals. This can be confirmed by a 1-NN classifier using the vectorized adjacency matrices, which gives an almost perfect classification accuracy. However, our goal here is to study the accuracy of low-dimensional representations of the graphs to discriminate between subjects. Thus, we measure how does our joint dimensional embedding method perform as the number of embedding dimensions grows. We compare the performance of our method with PCA, as another dimensionality reduction method. Note that the joint embedding of graphs directly imposes a rank-one restriction on the components, enforcing a low-rank representation of the graphs. PCA instead represents the adjacency matrices using components that are usually full rank, and requires many more parameters to represent those components.

Before computing the embedding, all graphs are centered by the mean, that is, we compute $\tilde\bA_i = \bA_i - \frac{1}{300}\sum_{j=1}^{300}\bA_j$. Using a 5-fold cross-validation, the data of each subject is divided into training and test sets. We measure the effect of the sample size using two different scenarios, the first uses $30$ graphs on the training set (one scan per subject) and the second uses $240$ graphs (eight scans per subject). The rest of the graphs are used in the test set to evaluate the accuracy, and the average over the 5 folds is computed. We follow the same procedure as in experiment 1. Using the training set, we compute an embedding of the graphs into $d$ dimensions, either by doing JEG or PCA on the vectorized adjacency matrices. After that, all the data is projected into the $d$-dimensional embedding, and we use 1-NN to estimate the labels of the test data. 

\begin{figure}[!htb]
	\centering
\includegraphics[width=0.5\textwidth]{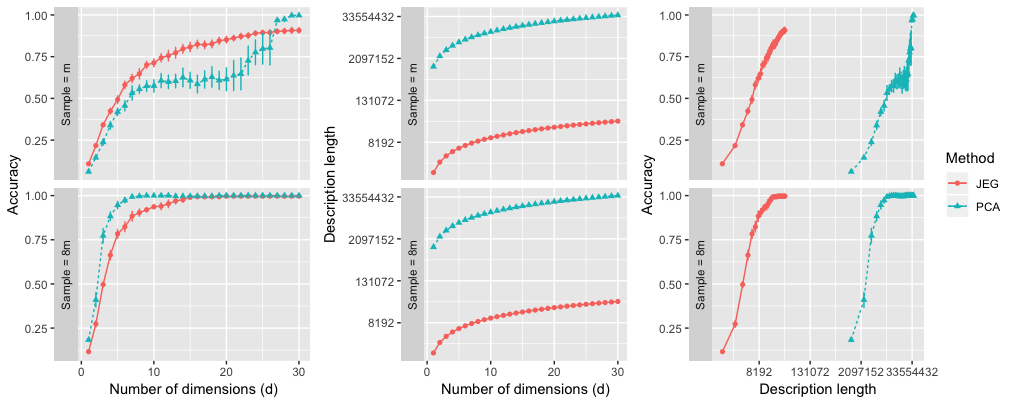}%
	\caption{Comparison of average classification accuracy and model complexity (number of embedding dimensions and description length) for JEG and PCA on the HNU1 data. A $d$ dimensional representation  is estimated using a training sample (top row: one graph per subject, bottom row: 8 graphs per subject), after which all data is embedded, and the test data is classified using 1-NN. The representations obtained by JEG are more accurate with fewer model parameters than PCA, especially when the sample size is small.}
	\label{fig:hnu1}
\end{figure}

\begin{figure}[!htb]
	\centering
\includegraphics[width=0.5\textwidth]{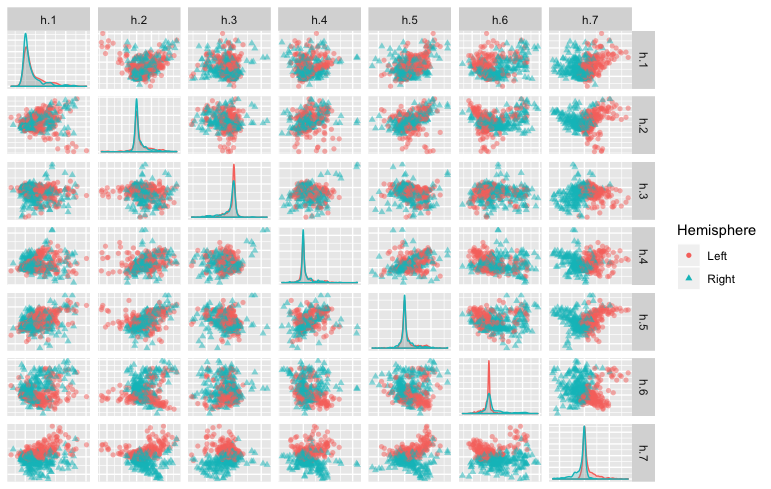}
	\caption{Latent positions of the vertices found by JEG for the first seven dimensions using the HNU1 data. The color indicates the hemisphere side according to the Talairach atlas. The off-diagonal panels contain scatter plots of the vertices and diagonal plots show the density. Several dimensions of the embedding show a relationship between the latent positions and the hemisphere side.}
	\label{fig:hnu1-eigenv}
\end{figure}

Figure \ref{fig:hnu1} shows a comparison of the average classification accuracy and model complexity for both methods. We measure model complexity as the number of embedding dimensions $d$, and as the total number of parameters used in each instance ($d(n+m)$ for JEG and $d(n^2+m)$ for PCA). In general, both methods are able to perform very accurate classification when $d$ is large enough, but JEG uses far fewer parameters in the components for the same number of dimensions (see middle plot). The advantage of JEG is especially remarkable when the sample size is small (first row). In this case, JEG shows a better performance than PCA when $d$ is not large. For values close to $30$, PCA shows almost perfect performance, which is not surprising, since the maximum number of principal components that can be obtained by PCA is $m=30$, and thus PCA is not really performing any dimensionality reduction when $d$ is close to this value, but rather projecting onto the training data itself. JEG on the other hand is able to provide very accurate classification error in all scenarios using low-rank representations of the graphs with a fewer number of parameters that are interpretable. These can be observed in Figure \ref{fig:hnu1-eigenv}, which shows the latent positions of the vertices obtained by JEG for the first seven dimensions. To construct these plots, only the vertices that are labeled as left (507 of them) or right (525) were considered. As it can be observed, several dimensions of the latent positions show a structure that is clearly related to the hemisphere side. This structure is specially highlighted on the 7th dimension of the embedding.

\subsection{Real Data Experiment 2: Joint Embedding to Cluster Vertices}
In the previous experiments, we focus on feature extraction for graphs. Here, we consider a different task, that is spectral clustering through the joint embedding. In general, spectral clustering first computes (generalized) eigenvalues and eigenvectors of
adjacency matrix or Laplacian matrix, then clusters the latent positions into groups \cite{sussman2012consistent, belkin2003laplacian}. The cluster identities of latent positions become the cluster identities of vertices of the original graph. Adjacency Spectral Embedding (ASE) is one of the spectral clustering algorithms used to find the latent positions of the vertices to fit the SBM and RDPG \cite{lyzinski2014perfect}, which are special cases of our MREG model. When applied to one graph, the joint embedding is equivalent to Adjacency
Spectral Embedding (ASE). When given multiple graphs, the joint embedding can estimate latent positions for graph $i$ as $[\hat{\lambda}_i[1]^{\frac{1}{2}}\hat{h}_1,\hat{\lambda}_i[2]^{\frac{1}{2}}\hat{h}_2,...,\hat{\lambda}_i[d]^{\frac{1}{2}}\hat{h}_d]$ or equivalently $\hat{\bH} \hat{\bD}_i ^ {\frac{1}{2}}$. Alternatively, the matrix of components $\hat{H} = [\hat{h}_1,\ldots,\hat{h}_d]$ gives joint latent positions for all the graphs. Then, a clustering algorithm can be applied to the latent positions to obtain communities.\\

We apply this spectral clustering approaches to two Wikipedia graphs \cite{suwan2016empirical}. The vertices of these graphs represent Wikipedia article pages. Two vertices are connected by an edge if either of the associated pages hyperlinks to the other. Two graphs are constructed based on English webpages and French webpages. The full graph has $1382$ vertices which represents articles within $2$-neighborhood of "Algebraic Geometry".  Based on the content of the associated articles, they are grouped by hand into $6$ categories: people, places, dates, things, math terms, and categories.\\

We consider a subset of vertices from $3$ categories: people, things, and math terms. After taking the induced subgraph of these vertices and removing isolated vertices, there are $n=704$ vertices left. Specifically, $326$, $181$, and $197$ vertices are from people, things and math terms respectively. We consider approaches to embed the graphs to obtain latent positions. First, we consider clustering of each graph separately by doing ASE on the English graph $\bA_{en}$ (ASE+EN), or equivalently, JE on $\bA_{en}$, and  ASE on the French Graph $\bA_{fr}$ (ASE+FR), and compare with the individual latent positions obtained by JE $\hat{\bH} \hat{\bD}_{en} ^ {\frac{1}{2}}$ (JE+EN) and $\hat{\bH} \hat{\bD}_{fr} ^ {\frac{1}{2}}$ (JE+FR). We also consider methods to estimate joint latent positions, by doing ASE on the mean of both graphs $\bar{\bA} = (\bA_{en} + \bA_{fr})/2$ (ASE+(EN+FR)) \cite{Tang2016-rc}, and the matrix $\hat{\bH}$ obtained by JE on both graphs (JE+(EN,FR)). The dimension $d$ is set to $3$ for all approaches, and the latent positions are scaled to have norm $1$ for degree correction. Then, we apply $3$-means algorithm to the latent positions. \\ 
\begin{table}[]
	\centering
	\begin{tabular}{|l| r r |}
		\hline
        Method & ARI & Purity \\\hline
        ASE+EN & 0.152 (0.02) & \textbf{0.555} (0.018)  \\
        JE+EN &  \textbf{0.154} (0.019) & 0.55 (0.019) \\\hline
        ASE+FR & 0.114 (0.019) & 0.521 (0.017)  \\
        JE+FR & \textbf{0.154} (0.019) & \textbf{0.549} (0.017) \\\hline
        ASE+(EN+FR) & 0.155 (0.02) & 0.547 (0.016) \\
        JE+(EN, FR) & \textbf{0.156} (0.02) & \textbf{0.551} (0.021)  \\\hline
	\end{tabular}
	\caption{Clustering Performance on Wikipedia Graphs. The adjusted rand index (ARI) and the purity of clustering  of different spectral clustering approaches are shown. The first two scenarios consider each graph separately, and the third considers them jointly. The standard error (included in parenthesis) is estimated through repeatedly clustering bootstrapped latent positions. The joint embedding estimates latent positions which combine the information in both graphs with good clustering performance.}
	\label{tb:wiki}
\end{table}

\begin{figure}[!htb]
	\centering
	\includegraphics[scale=0.9,width=3.5in]{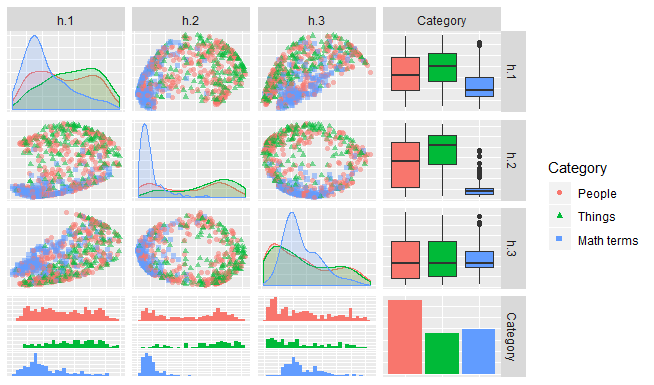}
	\caption{The joint latent positions of $\hat\bH$ estimated by the JE are shown. The first three plots on the diagonal are density estimates of latent positions for each dimension and category, and the last plot shows the number of points from each category. The first three plots of the last row show latent positions histograms for each dimension and category, and the first three plots of the last column are the corresponding box plots. The pairs plots of latent positions are given in the off-diagonal panels. The latent positions of math terms are separated from the other two clusters.}
	\label{fig:lp}
\end{figure}

The joint latent positions of the graphs $\hat\bH$ estimated by the joint embedding are provided on Figure \ref{fig:lp}. The latent positions of math terms are separated from the other two clusters. However, the latent positions of people and things are mixed. Table \ref{tb:wiki} shows the clustering results measured by adjusted rand index and the purity of clustering \cite{steinley2004properties,rendon2011internal}. The standard error is estimated through repeatedly clustering bootstrapped latent positions. All methods yield clustering results which are significantly better than random. The English graph demonstrates clearer community structure than the French graph. The joint embedding produces latent positions that combine the information in both graphs, and leads to better clustering performance. The JE and ASE give similar results on the English graph, but JE is able to improve the clustering performance on the French graph significantly. The JE also shows better performance than ASE on $\bar{\bA}$, which also uses the information of both graphs. These results show that our method provide  interpretable representations for the vertices, with an accuracy comparable to state of the art methods for spectral clustering.

\section{Conclusion}
In summary, we developed a joint embedding method that can simultaneously embed multiple graphs into low dimensional space. The joint embedding can be utilized to estimate features for inference problems on multiple vertex matched graphs. Learning on multiple graphs has significant applications in diverse fields and our results have both theoretical and practical implications for the problem. As the real data experiment illustrates, the joint embedding is a practically viable inference procedure. We also proposed a Multiple Random Eigen Graphs model. It can be understood as a generalization of the Random Dot Product Graph model or the Stochastic Block Model for multiple random graphs. We analyzed the performance of joint embedding on this model under simple settings. We demonstrated that the joint embedding method provides estimates with bounded error. Our approach is intimately related to other matrix and tensor factorization approaches such as singular value decomposition and CP decomposition. Indeed, the joint embedding and these algorithms all try to estimate a low dimensional representation of high dimensional objects through minimizing a reconstruction error. We are currently investigating the utility of joint embedding with more or less regularizations on parameters and under different set ups. We are optimistic that our method provides a viable tool for analyzing multiple graphs and can contribute to a deeper understanding of the joint structure of networks.

\section*{Appendix A}

\begin{proof} [Proof of Theorem 2.1]
	Denote the probability of observing a particular adjacency matrix $\bA_i$ under distribution $\mathcal{F}$ by $p_i$. It suffices to show that there is a set of parameters for MREG such that observing $\bA_i$ under MREG is also $p_i$. \\
	
	For undirected graphs with loops on $n$ vertices, there are ${n+1}\choose{2}$ possible edges. Let $\bA_1,\bA_2,...,\bA_{2^{{n+1}\choose{2}}}$ be all the possible adjacency matrices. 
	Since real symmetric matrix of size $n$ has ${n+1}\choose{2}$ free entries which lies in a linear space, if there exists ${n+1}\choose{2}$ linearly independent rank one symmetric matrices, they form a basis for this space. It turns out that the rank one symmetric matrices generated by vectors $\{e_i\}_{i=1}^{n} \cup \{e_i+e_j\}_{i<j}$ are linearly independent, where $\{e_i\}_{i=1}^{n}$ is the standard basis for $n$-dimensional Euclidean space. \\
	
	Next, we construct parameters for the MREG. Let $d$ be ${n+1}\choose{2} $  and 
	\[\{h_k\}_{k=1}^d = \{e_i\}_{i=1}^{n} \cup \{\frac{e_i+e_j}{\sqrt{2}}\}_{i<j}.\] 
	Since $\{h_k h_k^T\}_{k=1}^d$ forms a basis for real symmetric matrices, for each adjacency matrix $\bA_i$, there exists a vector $\lambda_i$, such that 
	\[\bA_i =\sum_{k} \lambda_i [k] h_k  h_k^T.\] Let $F$ be a finite mixture distribution on points $\{\lambda_i\}_{i=1}^{2^{{n+1}\choose{2}}}$, that is 
	\[F = \sum p_i \mathbb{I} \{\lambda = \lambda_i\}.\]
	Under this MREG model, for any adjacency matrix $\bA_i$
	\[P(\bA=\bA_i) = P(\lambda = \lambda_i) = p_i .\]
	This concludes that the distribution $\mathcal{F}$ and $MREG(F,h_1,...,h_d)$ are equal.
\end{proof}

\begin{proof}[Proof of Theorem \ref{theo:optimization}]
Suppose that $\hat{h}_{k_0} \hat{h}_{k_0}^T = \sum_{k=1}^{k_0-1} \alpha_k \hat{h}_k \hat{h}_k^T$. Therefore, from equation \eqref{eq:2} it can be noted that the solution for $\hat{\bLambda}_{*k_0}$ is given by
\begin{align*}
\hat{\bLambda}_{ik_0} & =   \underset{\bLambda_{ik_0}\in\Bbb{R}}{\operatorname{argmin}} \| \bA_i- \sum\limits_{k=1}^{k_0-1} \hat{\bLambda}_{ik} \hat{h}_{k} \hat{h}_{k}^T -\bLambda_{ik_0} h_{k_0} h_{k_0}^T\|  ^2.\\
& =   \underset{\bLambda_{ik_0}\in\Bbb{R}}{\operatorname{argmin}}  \| \bA_i- \sum\limits_{k=1}^{k_0-1} (\hat{\bLambda}_{ik} + \alpha_k \bLambda_{ik_0})\hat{h}_{k} \hat{h}_{k}^T \|  ^2.
\end{align*} 
Note that equation \eqref{eq:optimization-globalsolution} implies that for any $i=1,\ldots,m$,
\begin{equation*}
 \| \bA_i- \sum\limits_{k=1}^{k_0-1} (\hat{\bLambda}_{ik} + \alpha_k \bLambda_{ik_0})\hat{h}_{k} \hat{h}_{k}^T \|  ^2 \geq 
 \| \bA_i- \sum\limits_{k=1}^{k_0-1} \hat{\bLambda}_{ik}\hat{h}_{k} \hat{h}_{k}^T \|  ^2,
\end{equation*}
and hence, $\hat{\bLambda}_{ik_0}\sum\limits_{k=1}^{k_0-1} \alpha_k \hat{h}_{k} \hat{h}_{k}^T=0$, so $\hat{\bLambda}_{ik_0}=0$ for all $i=1,\ldots,m$. 
\end{proof}

\begin{proof} [Proof of Theorem 4.1]
	First,  we show that $|D_n(h,h_1)-D(h,h_1)|$ converges uniformly to $0$. To begin with, notice three facts:
	\begin{itemize}
		\item[(1)] the set $\{h: \|h\|=1\}$ is compact;
		\item[(2)] for all $h$, the function $\rho(\cdot,h)$ is continuous
		\item[(3)] for all $h$, the function $\rho(\cdot,h)$ is bounded by $n^2$.
	\end{itemize}
	Therefore, by the uniform law of large numbers \cite{jennrich1969asymptotic}, we have
	\[\underset{h}{\sup}|D_m(h,h_1)-D(h,h_1)|\overset{a.s.}{\rightarrow} 0.\]
	To prove the claim of the theorem, we use a technique similar to that employed by Bickel and Doksum \cite{bickel2015mathematical}. Let $h'$ be a global minimizer of $D(h, h_1)$. By definition, we must have $D_m(\hat{h}_1^m,h_1)  \leq D_m(h',h_1)$ and $D(h',h_1) \leq D(\hat{h}_1^m,h_1)$. From these two inequalities,
	\begin{align*}
	D_m(h',h_1)-D(h',h_1) &\geq D_m(\hat{h}_1^m,h_1)-D(h',h_1) \\
	&\geq D_m(\hat{h}_1^m,h_1)-D(\hat{h}_1^m,h_1) .
	\end{align*}
	Therefore, 
	\begin{align*}
	|D_m(\hat{h}_1^m,h_1)-D(h',h_1)|  \leq  \max(&|D_m(h',h_1)-D(h',h_1)|,\\
	& |D_m(\hat{h}_1^m,h_1)-D(\hat{h}_1^m,h_1)|).
	\end{align*}
	This implies 
	\[ |D_m(\hat{h}_1^m,h_1)-D(h',h_1)| \leq \underset{h}{\sup}|D_m(h,h_1)-D(h,h_1)|. \]
	Hence, $|D_m(\hat{h}_1^m,h_1)-D(h',h_1)|$ must converge almost surely to $0$, that is
	\[|D_m(\hat{h}_1^m,h_1)-D(h',h_1)|\overset{a.s.}{\rightarrow} 0 .\]
	If $\hat{h}_1^m$ does not converge almost surely to a global minimizer of $D(h,h_1)$,  then for all $h'\in\arg\min D(h,h_1)$, $\|\hat{h}_1^m-h'\|\geq \epsilon$ for some $\epsilon$ and infinitely many values of $m$. Since $h'$ is a global minimum, $|D(\hat{h}_1^m,h_1)-D(h',h_1)| > \epsilon' $ for infinitely many values of $m$ and some $\epsilon' $. This contradicts with the previous equation. Therefore, $\hat{h}_1^m$ must converge almost surely to $\arg\min D(h',h_1)$.
\end{proof}

\begin{proof} [Proof of Theorem 4.2]
	The proof of theorem relies on two lemmas. The first lemma shows that
	$h'$ is the eigenvector corresponding to the largest eigenvalue of $E(\langle \bA_{i},h' h'^T \rangle \bA_{i})$. The second lemma shows that $E(\langle \bA_{i},h' h'^T \rangle \bA_{i})$ is close to $E(\lambda_i^2) (h_1^Th')^2 h_1 h_1^T$ under Frobenius norm. Then, we apply Davis-Kahan theorem \cite{davis1970rotation} to establish the result of theorem.\\
	\begin{lemma}
		The vector $h'$ is the eigenvector corresponding to the largest eigenvalue of $E(\langle \bA_{i},h' h'^T \rangle \bA_{i})$.
	\end{lemma}
	We notice that
	\begin{align*}
	\underset{\|h\| =1}{\operatorname{min}}D(h,h_1) &=\underset{\|h\| =1}{\operatorname{min}}E(\|\bA_i- \langle \bA_i,h h^T \rangle h h^T\|^2) \\
	&=\underset{\|h\| =1}{\operatorname{min}}E(\langle \bA_i,\bA_i \rangle- \langle \bA_i,h h^T \rangle ^2) \\
	&=E(\langle \bA_i,\bA_i \rangle)-\underset{\|h\| =1}{\operatorname{max}}E( \langle \bA_i,h h^T \rangle ^2).
	\end{align*}
	Therefore, 
	\begin{equation} \label{eq:5}
	h'= \underset{\|h\| =1}{\operatorname{argmin}} \text{ }D(h,h_1)=\underset{\|h\| =1}{\operatorname{argmax}} \text{ } E(\langle \bA_i,h h^T \rangle ^2) .
	\end{equation}
	Taking the derivative of $E( \langle \bA_i,h h^T \rangle ^2)+ c(h^Th-1)$ with respect to $h$,
	\begin{align*}
	\frac{\partial E( \langle \bA_i,h h^T \rangle ^2)+ c(h^Th-1) }{\partial h} & =  E(\frac{\partial  \langle \bA_i,h h^T \rangle ^2}{\partial h}) +2ch \\
	&=4 E( \langle \bA_i,h h^T \rangle \bA_i)h +2ch .
	\end{align*}
	Setting this expression to $0$ yields,
	\begin{align*} 
	E( \langle \bA_i,h' h'^T \rangle \bA_i)h' & = -\frac{1}{2}ch' .
	\end{align*}
	Using the fact that $\|h'\|=1$, we can solve for $c$:
	\[c = -2 h'^T E( \langle \bA_i,h' h'^T \rangle \bA_i)h' = -2 E( \langle \bA_i,h' h'^T \rangle^2) .\]
	Then, substituting for $c$, 
	\begin{equation*}
	E( \langle \bA_i,h' h'^T \rangle \bA_i)h'=E( \langle \bA_i,h' h'^T \rangle^2)h'.
	\end{equation*}
	Therefore, we see that $h'$ is an eigenvector of $E(\langle \bA_{i},h' h'^T \rangle \bA_{i})$ and the corresponding eigenvalue is $E(\langle \bA_{i},h' h'^T \rangle ^2)$. Furthermore, $E(\langle \bA_{i},h' h'^T \rangle ^2)$ must be the eigenvalue with the largest magnitude. For if not, then there exists an $h''$ with norm $1$ such that
	\begin{align*}  
	| h''^T E(\langle \bA_{i},h' h'^T \rangle \bA_{i}) h''| &= |E(\langle \bA_{i},h' h'^T
	\rangle \langle \bA_{i},h'' h''^T \rangle)|\\
	& > E(\langle \bA_{i},h' h'^T \rangle ^2);
	\end{align*}
	however, by Cauchy-Schwarz inequality we must have
	\begin{align*}
	E(\langle \bA_{i},h'' h''^T \rangle^2) & E(\langle \bA_{i},h' h'^T \rangle^2)  \\ 
	& \geq \quad |E(\langle \bA_{i},h' h'^T \rangle \langle \bA_{i},h'' h''^T \rangle)|^2,
	\end{align*}
	implying $E(\langle \bA_{i},h'' h''^T \rangle^2) > E(\langle \bA_{i},h' h'^T \rangle^2)$, which contradicts equation \eqref{eq:5} of this appendix. Thus, we conclude that $h'$ is the eigenvector corresponding to the largest eigenvalue of $E(\langle \bA_{i},h' h'^T \rangle \bA_{i})$. \\
	\begin{lemma}
		\[	\|E(\langle \bA_{i},h' h'^T \rangle \bA_{i} ) - E(\lambda_i^2) (h_1^Th')^2 h_1 h_1^T\| \leq 2 E(\lambda_i).\]
	\end{lemma}	
	We compute $E(\langle \bA_{i},h' h'^T \rangle \bA_{i})$ by conditioning on $P_i$.
	\begin{align*}
	E(\langle \bA_{i},h' h'^T \rangle \bA_{i}|\bP_i)   
	&= && E( \langle \bA_{i}-\bP_i,h' h'^T \rangle (\bA_{i}-\bP_i)|\bP_i)\\
	& && +E(\langle \bA_{i}-\bP_i,h' h'^T \rangle \bP_i)|\bP_i) \\
	& && +E(\langle \bP_i,h' h'^T \rangle (\bA_{i}-\bP_i)|\bP_i)\\
	& && +E(\langle \bP_i,h' h'^T \rangle \bP_i|\bP_i) \\
	&= && E(\langle \bA_{i}-\bP_i,h' h'^T \rangle (\bA_{i}-\bP_i)|\bP_i)\\
	& && + \lambda_i (h_1^Th')^2 \bP_i \\
	&= && 2h' h'^T *\bP_i*(\bJ-\bP_i) \\
	& && - \operatorname{diag}(h_1 h_1^T *\bP_i*(\bJ-\bP_i))\\
	& && +\lambda_i (h_1^Th')^2 \bP_i .
	\end{align*}
	Here, $\operatorname{diag}()$ means only keep the diagonal of the matrix; $*$ means the Hadamard product, and $\bJ$ is a matrix of all ones. Using the fact that $\bP_i=\lambda_i h_1 h_1 ^T$, we have 
	\begin{align*}
	&E(\langle \bA_{i},h' h'^T \rangle \bA_{i}) - E(\lambda_i^2) (h_1^Th')^2 h_1 h_1^T \\
	&=E(E(\langle \bA_{i},h' h'^T \rangle \bA_{i}|\bP_i)-\lambda_i (h_1^Th')^2 \bP_i) 
	\\
	&= E(2 h' h'^T *\bP_i*(\bJ-\bP_i) - \operatorname{diag}(h' h'^T*\bP_i*(\bJ-\bP_i))).
	\end{align*}
	If we consider the norm difference between $E( \langle \bA_{i},h' h'^T \rangle \bA_{i})$ and $ E(\lambda_i^2) (h_1^Th')^2 h_1 h_1^T$, we have
	\begin{align*}
	&\|E(\langle \bA_{i},h' h'^T \rangle \bA_{i} ) - E(\lambda_i^2) (h_1^Th')^2 h_1 h_1^T\| \\
	&= \|E(2 h' h'^T *\bP_i*(\bJ-\bP_i)  - \operatorname{diag}(h' h'^T*\bP_i*(\bJ-\bP_i)))\| \\
	&\leq  E(\|2 h' h'^T *\bP_i*(\bJ-\bP_i)  - \operatorname{diag}(h' h'^T*\bP_i*(\bJ-\bP_i))\|) \\
	&\leq  E(\|2 h' h'^T *\bP_i*(\bJ-\bP_i)\|) \\
	&\leq  E(\|2 h' h'^T * \bP_i\|) \\
	&\leq  2 E(\lambda_i)\| h' h'^T * h_1 h_1^T\| \\
	&=  2 E(\lambda_i) .
	\end{align*}
	This finishes the proof for the lemma. \\
	
	Notice that the only non-zero eigenvector of $E(\lambda_i^2) (h_1^Th')^2 h_1 h_1^T$ is $h_1$ and the corresponding eigenvalue is $E(\lambda_i^2) (h_1^Th')^2$. We apply the Davis-Kahan theorem \cite{davis1970rotation} to the eigenvector corresponding to the largest eigenvalue of matrices $E(\langle \bA_{i},h' h'^T \rangle \bA_{i} )$ and $E(\lambda_i^2) (h_1^Th')^2 h_1 h_1^T$, yielding
	\[\|h'-h_1\| \leq \frac{2 E(\lambda_i)}{E(\lambda_i^2)(h_1^T h')^2}. \]
	
\end{proof}
 
\bibliographystyle{ieeetr}
\bibliography{mybibfile}


%

\vspace{-10 mm}
\begin{IEEEbiography}[{\includegraphics[width=1in,height=1.25in,clip,keepaspectratio]{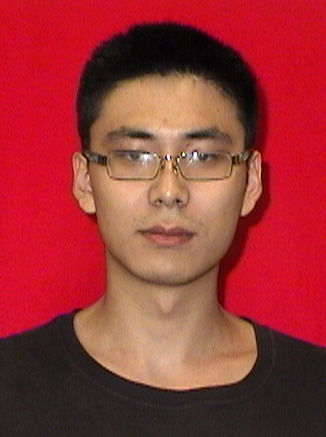}}]{Shangsi Wang}
	received the BS degree in mathematics and actuarial science from	
	University of Waterloo in 2012, the PhD degree from the Department of Applied Mathematics and Statistics at Johns Hopkins University in 2019. He is currently working in the financial services industry. His research interests include statistical inference on random netowrks and pattern recognition in graph datasets.
\end{IEEEbiography}

\begin{IEEEbiography}[{\includegraphics[width=1in,height=1.25in,clip,keepaspectratio]{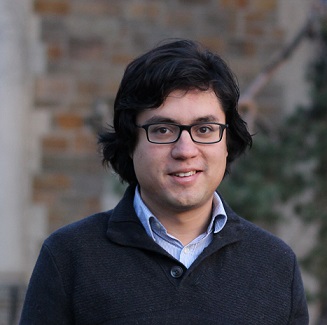}}]{Jes\'us Arroyo}
	received the BS degrees in applied mathematics and computer engineering from the Instituto Tecnol\'ogico Aut\'onomo de M\'exico
	(ITAM) in 2013, and the MA and PhD degrees from the Department of Statistics at the University of Michigan, Ann Arbor, in 2018. He is postdoctoral fellow in the Center for Imaging Science at Johns Hopkins University since 2018. His research interests include statistical analysis of graphs and high dimensional data, and applications to neuroimaging.
\end{IEEEbiography}

\begin{IEEEbiography}[{\includegraphics[width=1in,height=1.25in,clip,keepaspectratio]{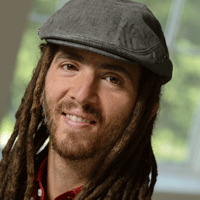}}]{Joshua T. Vogelstein}
	received the BS degree from the Department of Biomedical Engineering
	(BME) at Washington University in St. Louis, in 2002, the MS degree from the Department of Applied Mathematics and Statistics at Johns Hopkins University (JHU), in 2009, and the PhD degree from the Department of Neuroscience at JHU in 2009. He was a postdoctoral fellow in AMS at JHU from 2009 until 2011, when he was appointed an assistant research scientist, and became a member of the Institute for Data Intensive Science and Engineering. He spent two years at Information Initiative at Duke, before becoming Assistant Professor in BME at JHU, and core faculty in the Institute for Computational Medicine and the Center for Imaging Science. His research interests include computational statistics, focusing on ultrahigh-dimensional and non-Euclidean neuroscience data, especially connectomics.
\end{IEEEbiography}

\begin{IEEEbiography}[{\includegraphics[width=1in,height=1.25in,clip,keepaspectratio]{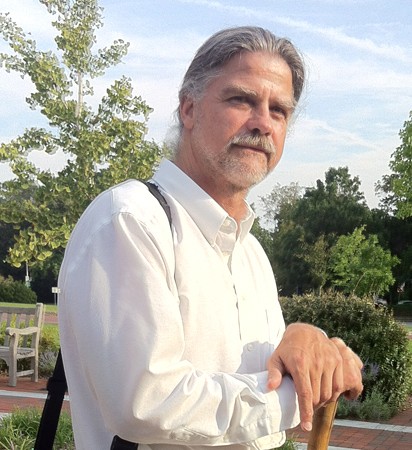}}]{Carey E. Priebe}
	received the BS degree in mathematics from Purdue University, in 1984, the MS degree in computer science from San Diego State University, in 1988, and the PhD 	degree in information technology (computational	statistics) from George Mason University, in 1993. From 1985 to 1994, he worked as a mathematician and scientist in the US Navy research and development laboratory system. Since 1994, he has been a professor in the Department of Applied Mathematics and Statistics, Johns Hopkins University (JHU). His research interests include computational statistics, kernel and mixture estimates, statistical pattern recognition, statistical image analysis, dimensionality reduction, model selection, and statistical inference for high-dimensional and graph data. He is a Senior Member of the IEEE, a Lifetime Member of the Institute of Mathematical Statistics, an Elected Member of the International Statistical Institute, and a Fellow of the American Statistical Association.
\end{IEEEbiography}

%


\end{document}